\documentclass[prb,twocolumn,showpacs,amsmath,amssymb,superscriptaddress,preprintnumbers,floatfix]{revtex4}

\usepackage{graphicx}
\newcommand{\bA}{{\bf A}}
\newcommand{\ba}{{\bf a}}
\newcommand{\bB}{{\bf B}}
\newcommand{\bG}{{\bf G}}

\newcommand{\bk}{{\bf k}}
\newcommand{\bl}{{\bf l}}

\newcommand{\bp}{{\bf p}}

\newcommand{\bR}{{\bf R}}
\newcommand{\br}{{\bf r}}

\newcommand{\bv}{{\bf v}}
\newcommand{\bw}{{\bf w}}

\newcommand{\hx}{\hat{x}}
\newcommand{\hy}{\hat{y}}

\newcommand{\cA}{{\cal A}}
\newcommand{\cE}{{\cal E}}

\newcommand{\cV}{{\cal V}}
\newcommand{\eps}{\epsilon}

\newcommand{\tmpQ}{k_D}
\newcommand{\tmpDelta}{\Delta}
\newcommand{\bdelta}{\boldsymbol{\delta}}
\newcommand{\sign}{\mathop{\rm sgn}\nolimits}
\begin{document}
\bibliographystyle{apsrev}

\title{Mixed state of a lattice $d$-wave superconductor}

\author{Ashot Melikyan}%
\email{ashot@phys.ufl.edu}
\affiliation{Institute of Fundamental Physics, Department of Physics, University of Florida, Gainesville, FL, 32611}
\author{Zlatko Te\v{s}anovi\'{c}}%
\email{zbt@pha.jhu.edu}
\affiliation{Department of Physics and Astronomy, Johns Hopkins University, Baltimore, MD, 21218}

\date{\today}
\begin{abstract}
We study  the mixed state in an extreme 
type-II lattice $d_{x^2-y^2}$-wave superconductor
in the experimentally most relevant
regime of intermediate magnetic fields $H_{c1}\ll H \ll H_{c2}$. 
We analyze the low energy
spectrum of the problem dominated by  nodal Dirac-like 
quasiparticles with momenta near ${\bf k}_F=(\pm k_D,\pm k_D)$ and find
that the spectrum exhibits characteristic 
oscillatory behavior with respect to the
product of $k_D$ and magnetic length $l$. The Simon-Lee scaling,
predicted in this regime, is satisfied only on average, with the magnitude
of the oscillatory part of the spectrum displaying the same $l^{-1}$
dependence as its monotonous ``envelope'' part. In general, the
spectrum obeys a scaling law 
$E_{n {\bf k}} = \displaystyle\frac{\hbar v_F}{l} \cE_n({\bf k} l, t/\Delta, k_D l)$,
where ${\cal E}$ is a 
dimensionless universal $2\pi$-periodic function of $k_D l$.
The oscillatory behavior of the spectrum is due to the inter-nodal
interference enhanced by the singular nature of the low energy
eigenfunctions near vortices. 
Our results constitute an example of a finite size scaling
of the Dirac-type quantum criticality.
We also study a separate problem of a single vortex piercing an
isolated superconducting grain of size $L\times L$. Here we find
that the periodicity of the quasiparticle
energy oscillations with respect to $k_D L$
is doubled relative to the case where the field is zero
and the vortex is absent, both such oscillatory behaviors being
present at the leading order in $L^{-1}$.
Finally, we review the overall features of the tunneling conductance
experiments in YBCO and
BSCCO,
and suggest an
interpretation of
the peaks at $5-20$ meV  observed in the tunneling
local density of states in these materials. 
We find that in the case of a pure $d$-wave superconducting order parameter
with featureless vortex cores,
the zero bias conductance peak (ZBCP) appears
only on the sites that are the immediate nearest neighbors
of vortex locations,
while all the other sites in the close
vicinity of vortices exhibit no such ZBCP, and
instead display pronounced peaks at sub-gap energies, typically
at about a half or less of the coherence peak energy.
Furthermore, we find that the on-site ZBCP can be 
strongly suppressed by enhanced local pairing near a vortex.
\end{abstract}

\pacs{74.25.Jb, 74.25.Qt}

\maketitle
\section{Introduction}
\label{section:introduction}

The study of the quasiparticle spectrum in
the mixed state of $d$-wave superconductors followed soon \cite{volovik,
soininen, wang_macdonald}
after the $d$-wave nature of pairing in the cuprate superconductors became
apparent. A distinct character of the 
Tunneling Local Density of States (TLDOS) 
in $s$-wave and $d$-wave superconductors near a 
vortex  was proposed by Wang and MacDonald to serve as a test which
would allow determination of the pairing symmetry in the cuprates \cite{wang}.
For $d$-wave pairing, they found that the TLDOS 
at the vortex location plotted as a function of
energy,  exhibits a prominent peak at zero applied bias voltage, while 
in an $s$-wave case, for otherwise similar parameters of the model, 
the thermally broadened TLDOS has a minimum at $E=0$ surrounded by two large
sub-gap peaks. While their calculations apply to densely packed
vortices and unrealistically high
magnetic fields in real cuprates, a similar conclusion concerning
this ``zero-bias peak'' is obtained also within
a single vortex calculation of Franz and Te\v{s}anovi\'{c} \cite{ft_DPLUSID}.
Interestingly, the STM experiments in BSCCO\cite{pan}
and YBCO \cite{maggio-aprile} -- now unambiguously known to be of a $d$-wave
type -- reveal that the ``zero-bias peak'' is completely absent.
Instead, the tunneling conductance experiences a dip at zero bias, and new
relatively small sub-gap peaks at energies $5-20$ meV.
To explain this discrepancy -- which at the moment is
unresolved -- one usually relies on additional order 
parameters in the vortex cores. This line 
of thought advocates that due to the
suppression of the superconducting order parameter within vortex cores, a
new competing (local) order emerges there, 
which ultimately is responsible for the
deviations from Refs. \onlinecite{wang,ft_DPLUSID}.
Several different order parameters were considered in the
literature:  $d+id$ superconducting order\cite{ft_DPLUSID},
antiferromagnetic order \cite{arovas}, pseudogap state \cite{ft_MOTT}, 
circulating currents \cite{han,kishine}, d-density-wave \cite{nayak}. Other
explanations of the absent ZBCP include  the
anisotropy of the tunneling matrix elements \cite{wu}, and, most recently,
quantum zero-point motion of vortices\cite{sachdev_QZPM}.

In addition to the above ``high-energy, short-distance'' 
features of a vortex core
-- an interesting problem in its own right --
the quest for a description of nodal quasiparticles
within some simple low-energy effective
Hamiltonian, which would facilitate theoretical analysis,
was launched as a separate line of inquiry. 
An effective description is important
in order to address more complicated problems
such as the interactions of fluctuating vortices
with nodal $d$-wave 
quasiparticles or the effects of disorder in a nodal
superconductor.
The initial steps in this direction were taken
by Simon and Lee \cite{simonlee} who proposed that, after extracting
the rapid ``$k_F$'' oscillations of the wavefunctions,  the ``linearized''
effective version of the Bogoliubov-deGennes (BdG) Hamiltonian
suggests a simple scaling (\ref{simonleescaling}) for the 
low energy ($E\ll \Delta$) sector of the
quasiparticle spectrum  in the mixed state
of type II-superconductors, and
consequently for various other measurable quantities. 
The scaling function was then calculated by Franz
and Te\v{s}anovi\'{c}\cite{ft_BLOCH}, who employed a 
singular gauge transformation
(FT transformation) and expansion of the wavefunctions 
in the plane wave basis to
find that the spectrum at the very low energies $E\ll \hbar v_F/l$ is 
essentially the same as the original spectrum of the zero-field  problem,
but for the renormalization of the slopes of the anisotropic Dirac cones at
the nodes. The linearized FT Hamiltonian was 
subsequently analyzed both numerically and
theoretically, using its symmetry 
properties, in Refs. \onlinecite{mhs,vishwanath_prb}.

Further study, however, revealed several new questions. 
Quite separately from its origin, the linearized Hamiltonian turned out
to be somewhat challenging to analyze due to singularities at vortex
positions which
rendered it incomplete unless its proper self-adjoint 
extension is constructed by imposing an additional 
boundary condition at
each vortex\cite{vmft,vmt}. Such boundary conditions, which turned out 
to be necessary in performing the numerical and
symmetry analysis of the linearized Hamiltonian, are
discussed at length in a companion paper\cite{mt_SAE}.

Furthermore, the relation of the linearized description to the tight-binding
model also turned out to be somewhat more complex than 
initially anticipated: the Simon-Lee scaling of the quasiparticles 
energy eigenstates according to
(\ref{simonleescaling}) demands that if the spectrum is gapless on the
linearized level -- as found in Ref. \onlinecite{ft_BLOCH} -- the
gaps of the full  non-linearized problem 
must  decrease as $1/l^2$ as a 
function of $l$ or faster in the limit of small magnetic fields.
However, exact diagonalization  of the non-linearized problem at
zero chemical potential $\mu=0$,
showed\cite{vmft} that the gap in the spectrum oscillates
between $0$ and ${\cal O}(\hbar v_F/l)$, depending on the commensuration
of the magnetic length to the atomic
lattice spacing. Perhaps the most
telling manifestation of the intricate relation 
between the linearized continuum
and the tight-binding lattice models of the
mixed state is
the exact result\cite{vm_index} for 
the spectrum of the latter when
$\mu=0$ and $l/\delta=2\pmod{4}$: in this case
the number of the zero energy states is
doubled compared to the zero magnetic field result. Clearly, such doubling is
difficult to account for if one uses the non-perturbed
plane wave basis as the departure point for a perturbation theory.

Here, we explore further the non-perturbative effects of the
tight-binding (TB) model of a $d$-wave superconductor in the presence of a
vortex lattice. In section II we
present the results of a systematic study of the spectrum for 
large magnetic lengths $l$ (low magnetic fields, corresponding
to realistic values in cuprates), up to $l=120\delta$,
where $\delta$ is the lattice spacing, and for general $\mu$. We
start by focusing on the {\em low-energy properties of the spectrum} and
analyze the validity Simon-Lee scaling for this model. 
The dispersion is
shown to obey the scaling on
{\em average}, and in general to experience rapid
oscillations of the energy levels as a
function of both the magnetic field and $\mu$. These 
anomalous oscillations, which can be unified
within a new generalized form of Simon-Lee scaling, are described by an
additional dependence of the energy levels on 
the commensuration of the inter-nodal distance and 
magnetic length $l$.
The inadequacy of the Simon-Lee scaling in its conventional form
is shown to be the result of the
singular nature of the BdG eigenfunctions combined 
with the inter-nodal interference, as conjectured 
in Refs. \onlinecite{vmft,vmt}. The linearized effective Hamiltonian is
argued to still accurately represent the low energy sector of the theory,
but the necessary condition is stricter than anticipated earlier and 
demands also $k_F \xi \gg 1$ rather than only $k_F l \gg 1$.

In section III we describe the {\em high-energy, short-distance
features of the spectrum}. We find that
although the TLDOS is indeed peaked at the four sites of
the tight-binding lattice surrounding the vortex, in agreement with previous
work, the four immediate neighbors are rather an exception than a rule: all
sites in the proximity of the vortex -- {\em except} the 
nearest neighbors -- exhibit
no zero-bias peak, and furthermore have additional peaks at sub-gap energies.
In the concluding section, we discuss 
how the on-site peak can be suppressed and
argue that the $5-7$ meV peaks observed in STM
experiments  could in fact be due to regular $d$-wave
vortices, but with a particular profile for the amplitude of 
$d$-wave gap function on those few bonds constituting the cores.

The anomalous enhancement
of the inter-nodal interference by singular potential due to vortices has
also a prominent effect on a related single vortex problem,
i.e., an isolated superconducting
grain of size $L\times L$ in a commensurate magnetic field $H = \Phi_0
L^{-2}$, where the elementary flux $\Phi_0$ equals $hc/(2e)$, pierced by a
single vortex. This is discussed in detail in section IV, where we show that
although the spectrum has oscillations of the low energy levels of magnitude
proportional  $L^{-1}$ even in  zero magnetic field due to finite-size effects,
in the presence of the vortex the periodicity of these oscillations is doubled.
\section{Tight-binding lattice model}
\label{section:tightbinding}

\subsection{BdG equations}
\label{section:BdG_equations}
We start from the Bogoliubov-deGennes(BdG) Hamiltonian $H_{TB}$ of the model
\cite{vmft,vmt}, which is defined by its action on a two-component Bogoliubov-Nambu wavefunction
$\psi_{\br} =(u_{\br}, v_{\br})^T$ as
\begin{equation}
H_{TB}\psi_{\br}=
\sum\limits_{\br'}
\begin{pmatrix}
t_{\br\br'} -\mu\delta_{\br\br'} & \Delta_{\br\br'}\\
\Delta^*_{\br'\br} & \mu\delta_{\br\br'} - t^*_{\br\br'}
\end{pmatrix}\psi_{\br'}.
\label{TB_Ham_start}
\end{equation}
In the simplest case, the hopping and pairing fields  described by $t_{\br\br'}$
and $\Delta_{\br\br'}$ are nonzero only on the nearest neighbor
bonds, and in the presence of magnetic field $\bB$,
the hopping $t_{\br\br'}=t^*_{\br'\br}$ is modified by Peierls factors
\begin{equation}
t_{\br\br'} =-t \exp(-i A_{\br\br'})\equiv -t
\exp\left(-\frac{ie}{ \hbar c}\int_{\br}^{\br'} \bA \cdot d\bl\right),
\end{equation}
where $\bA$ is the vector potential corresponding to magnetic field $\bB$.
The pairing field/gap function
 $\Delta_{\br\br'}$ should in principle be
determined from a self-consistent procedure stemming from the same
microscopic Hamiltonian that in the 
mean-field approximation yielded (\ref{TB_Ham_start}). For
example, in the simplest model that results in the $d$-wave order  within the
mean-field approximation for a wide
range of parameters -- the extended Hubbard model with the nearest
neighbors density-density attraction $g n_{\br} n_{\br'}$ -- the 
self-consistency condition reads
\begin{equation}
\Delta_{\br\br'} = g
\sum_{n} (
u_{\br } v^*_{\br'}+
u_{\br'} v^*_{\br}
)
\tanh \frac{E_n}{2T}~~,
\label{TB_selfconsistency}
\end{equation}
where $T$ is the temperature and $(u_n, v_n)$ are the eigenstates of the BdG
Hamiltonian of energy $E_n$.
While incorporating the self-consistency condition is not an impossible
task, the results to a certain extent will depend on the
microscopic model, from which the condition was derived.

In the context of the superconductivity in cuprates, however,
such dependence is very weak: the amplitude of the order parameter
$|\Delta_{\br\br'}|$ recovers rapidly to its uniform state value
$\tmpDelta$, while the phase is subject to the condition of overall winding
by $2\pi$ along any lattice path enclosing a vortex. These two conditions
suggest a useful simple Ansatz for the starting point of the iterative
self-consistency procedure: $\Delta_{\br\br'} = \tmpDelta \eta_{\br\br'}
\exp(i\theta_{\br\br'})$, where the bond phase
$\theta_{\br\br'}$ is given in the Appendix
\ref{appendix_tightphases}
and the $d$-wave nature of the bond field $\Delta_{\br\br'}$
enters through factors $\eta_{\br,\br+\bdelta}=1$ 
($\eta_{\br,\br+\bdelta}=-1$) 
if $\bdelta=\pm \delta\hat{x}$ ($\bdelta=\pm \delta \hat{y}$).
One then proceeds to diagonalize $H_{TB}$ from (\ref{TB_Ham_start}), 
recomputes $\Delta_{\br\br'}$
using the self-consistency condition (\ref{TB_selfconsistency}),
and repeats the procedure until the convergence is achieved. In practice, the 
starting Ansatz is a very good approximation to the final
solution in a sense that both have the same phase defects, the same
symmetries, and the ratio of  the fully self-consistent solution
$\Delta_{\br\br'}$ to the Ansatz $\tmpDelta\eta_{\br\br'}\exp(i\theta_{\br\br'})$ is
merely a periodic smooth function close to unity at all bonds of the 
lattice, except possibly in a close proximity of vortices.
Consequently, we first concentrate on the 
(non-selfconsistent) BdG Hamiltonian 
 (\ref{TB_Ham_start}) with $\Delta_{\br\br'} = \tmpDelta \eta_{\br\br'}
\exp(i\theta_{\br\br'})$, which allows for an easier systematization of the
results due to the reduced number of parameters and note that 
all possible microscopic Hamiltonians leading to $d$-wave pairing
are now encoded by a single bond variable $\tmpDelta$.
At the same time, this approach permits one to avoid a time-consuming
search for the solution of the fully 
self-consistent problem. In the end, we will briefly
discuss the effects of varying $\Delta_{\br\br'}$ near vortices.

\begin{equation}
H_{TB}\psi_{\br} =
\sum_{\br'}
\begin{pmatrix}
-t
e^{-i{A}_{\br\br'}}&
\tmpDelta
\hat{\eta}_{\br\br'}
e^{i\theta_{\br\br'}}\\
\tmpDelta 
\hat{\eta}_{\br\br'}
e^{-i\theta_{\br'\br}}&
te^{i{A}_{\br\br'}}
\end{pmatrix}\psi_{\br'}-\mu\sigma_3\psi_{\br},
\label{TB_Ham}
\end{equation}
where the summation indices $\br'$ denote the nearest neighbors of $\br$ and
$\sigma_3$ is the Pauli matrix.
Although we will be interested in a periodic inversion-symmetric lattice of
vortices, Hamiltonian $H_{TB}$ does not 
explicitly possess the symmetries of the
physical vortex lattice. In general, one has to accompany
the translations by a vortex lattice vector with an additional gauge
transformation restoring the Hamiltonian to its
original form. Rather than working with
representations of the resulting magnetic translations group, we perform a
unitary transformation of a special form  
$U=\mbox{diag}(e^{i\alpha_{\br}}, e^{-i\beta_{\br}})$
such that the transformed Hamiltonian 
$H=U^{-1}H_{TB}U$ is explicitly periodic.

It is easy to see that, regardless of the transformation used to bring the
Hamiltonian  $H_{TB}$ to a
periodic form, the minimal unit cell must contain at least two
vortices: after the transformation the diagonal hopping term contains 
modified Peierls factors
$\exp(i\tilde{A}_{\br\br'})=\exp(i{A}_{\br\br'}+i\alpha_{\br'}-i\alpha_{\br})$.
Suppose these factors are periodic, then 
consider  a product $\prod_{\langle\br\br'\rangle} e^{i\tilde{A}_{\br\br'}}$
over the oriented bonds along a closed path formed by 
the  perimeter of the unit cell
traversed counterclockwise. Since $e^{i{A}_{\br\br'}} = e^{-i{A}_{\br'\br}}$
and factors  $\exp(i\tilde{A}_{\br\br'})$ are to be periodic, such a product
should be equal to $1$.
$$
\prod_{\langle\br\br'\rangle} e^{i\tilde{A}_{\br\br'}}=
\prod_{\langle\br\br'\rangle} e^{i{A}_{\br\br'}}  = e^{\displaystyle-\oint
\frac{ie}{\hbar c}\bA \cdot
d\bl} = 1.
$$
Thus the flux of magnetic field through the unit cell must be an integer of
$2\pi\hbar c/e = hc/e$,  and therefore must contain at least two 
$hc/2e$ superconducting vortices.

\begin{figure}[tbh]
\centering
\includegraphics[width=0.7\columnwidth]{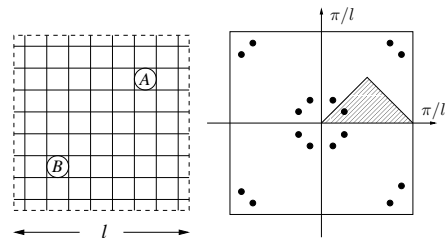}

\caption{\label{unitcell} Left: magnetic unit cell containing two
vortices labelled as $A$ and $B$ with magnetic length $l=8\delta$. Right:
The symmetry of dispersion $E_{n\bk}$ within the Brillouin zone (BZ), which 
follow from the symmetry operations of the Hamiltonian $H$, is shown.
16 equivalent points are displayed as solid dots; and 
it is sufficient to study only $1/16$-th portion of the BZ drawn 
as a dashed triangle.
}
\end{figure}

An explicit form of such a transformation can be realized by 
considering a simple family of the so-called symmetric
transformations\cite{gorkov_schrieffer, anderson}:
$$
U_{{\mathbb Z}2}=\mbox{diag}(e^{i\alpha_{\br}},
e^{-i\alpha_{\br}}),
$$
with suitably chosen $\alpha_{\br}$.
In the continuum version of the theory this 
transformation forces branch-cuts and non-locality
on the eigenfunctions of
$H_{TB}'$. Although the tight-binding lattice version of it does 
not cause undue complications
\cite{vm_index}, here we utilize another common
choice, the FT transformation\cite{ft_BLOCH},
whose continuum analogue does not require branch-cuts:
\begin{equation}
U=\mbox{diag}(e^{i\phi^A_{\br}},
e^{-i\phi^B_{\br}}),
\end{equation}
where site variables $\phi_A$ and $\phi_B$ are 
explicitly given in the Appendix A of Ref. \onlinecite{mt_SAE}. 
After defining periodic\cite{vmft, mt_SAE}  
bond variables ${\cal V}^{A}_{\br\br'}$,  
${\cal V}^{B}_{\br\br'}$, and
${\cA}_{\br\br'}$ according to 
\begin{align}
\cV^{A(B)}_{\br\br'} &= \phi^{A(B)}_{\br'}-\phi^{A(B)}_{\br}-A_{\br\br'}\\
\cA_{\br\br'}& = \theta_{\br\br'}-\phi^A_{\br}-\phi^B_{\br'},
\label{definebondvars}
\end{align}
 the resulting periodic Hamiltonian is given by
\begin{equation}
H\psi_{\br} =
\sum_{\br'}
\begin{pmatrix}
-t
e^{i {\cal V}^A_{\br\br'}}&
\tmpDelta 
\hat{\eta}_{\br\br'}
e^{i{\cA}_{\br\br'}}\\
\tmpDelta 
\hat{\eta}_{\br\br'}
e^{-i{\cA}_{\br'\br}}u_{\br'}&
te^{-i\cV^B_{\br\br'}}
\end{pmatrix}\psi_{\br'}
-\mu \sigma_3\psi_{\br}
\label{TB_Ham_FT}
\end{equation}
with $\br'$ denoting the nearest neighbors of $\br$.
\subsection{Linearization of the tight-binding Hamiltonian}
\label{section:linearization}
One can now
attempt to describe the low energy portion of the quasiparticle spectrum
using the linearized approximation, leading to Simon-Lee scaling for various
properties.  The derivation of the linearized 
version of $H_{TB}$ has been performed in
Refs. \onlinecite{vmft,vmt} by gradient expansion 
resulting in a continuum Hamiltonian
with dispersion, which is then linearized as usual \cite{ft_BLOCH}. One
might expect that the coefficients of the final model obtained in this way
reproduce the spectrum of the full TB model for the values of chemical
potential $\mu$ not too close to the half-filling -- where the dispersion is
not quadratic -- and also for $\mu$ not too close to the
bottom of the tight-binding band, otherwise the linearization procedure itself
is not justified. Below, we describe an alternative
derivation which, while proceeding along similar lines
as the standard procedure, does lift the first restriction and allows one to
consider values of $\mu$ at and near half-filling. At the end of this
section, we will revisit this derivation and
show that it should be corrected to accommodate the curvature and the
so-called ``interference'' effects. 

The linearization procedure is based on
the assumption that wavefunctions can be represented as
\begin{equation}
\psi_{\br} = \sum_{j=1,\overline{1},2,\overline{2}} e^{i\bk_{F}^{(j)}\cdot \br} \psi^{(j)}(\br)~~,
\label{linearizedWF}
\end{equation}
where $j$ labels the Dirac-like nodes of a $d$-wave gap function located in
momentum space at $\bk_{F}^{(j)}=(\pm\tmpQ, \pm\tmpQ)$, and $\psi^{(j)}(\br)$ is slowly varying function
on the scale of $k_F^{-1}$. Variable $k_D$ introduced here for brevity of
notation  simply equals $k_F/\sqrt{2}$, where $k_F$ is the magnitude of
the Fermi momentum in a nodal direction.
After substituting this form into the BdG eigenvalue problem for Hamiltonian 
$H$ (\ref{TB_Ham_FT}), a typical term
has the form  of a sum  over $\bdelta=\pm \delta\hx, \pm \delta\hy$:
\begin{equation}
S=\sum_{\bdelta} e^{i\int_{\br}^{\br+\bdelta} \bw \cdot d\bl} f_{\br+\bdelta}~~,
\end{equation}
where Fourier transform of $f_{\br}$ is assumed to be composed of wavevectors
close to the four nodal momenta at the
Fermi surface and $\nabla\cdot \bw=0$. Note that the straightforward  gradient
expansion is valid only qualitatively since  $k_F \delta$ is 
not necessarily much
smaller than one. The linearization, however, can be performed directly,
without the preliminary ``continuization'' step, by first
separating the rapidly oscillating part: 
$$
f_{\br} = e^{i\bk_F^{(j)} \cdot\br} F_{\br}~~,
$$
where $F_{\br}$ is a function that changes slowly on the scale of few
lattice spacings. To obtain the effective linearized  description,
we now replace the lattice function $F_{\br}$
by a slowly changing function $F(\br)$ defined in
continuum, such that it coincides with $F_{\br}$ when $\br$ correspond
to the lattice sites. Thus, $F(\br)$ can be thought of as an
interpolating function for a discrete set $F_{\br}$. The detailed form of
the interpolation turns out unimportant for the leading order results
derived below, as long as the characteristic
scale on which $F_{\br}$ varies is much larger than the lattice
spacing $\delta$. Then, in the expression
$$
S = e^{i\bk_F^{(j)}\cdot \br}
\sum_{\bdelta} e^{i\int_{\br}^{\br+\bdelta} \bw \cdot d\bl}
e^{i \bk_F^{(j)}\cdot\bdelta} F(\br+\bdelta)~~,
$$
we use
$$
e^{i\int_{\br}^{\br\pm \delta\hx} \bw \cdot d\bl} =
1 \pm i w_x \delta -\frac{w_x^2 \delta^2}{2} +\frac{i\delta^2}{2}\nabla_x w_x~~.
$$
and expand slowly varying function $F(\br)$ into Taylor series 
while retaining factors $\exp({i \bk_F^{(j)}\cdot\bdelta})$. 
The result for node
$\bk_F^{(1)} =({\tmpQ},{\tmpQ})$ is
\begin{multline}
S= -e^{i\bk_F^{(1)} \br} 
\left[
2\sqrt{2} \delta \sin({\tmpQ}\delta)
\left(
\frac{p_x+p_y}{\sqrt{2}}+\frac{w_x+w_y}{\sqrt{2}}
\right)F \right.\\
\left.
+
\delta^2\cos({\tmpQ}\delta)
(\bp+\bw)^2 F+{\ldots} 
\right],
\end{multline}
where $\bp$ denotes the usual continuum momentum operator
$-i\nabla$, and $\ldots$ denote terms containing higher powers of
$\delta\nabla$ and $\delta\bw$.

The expressions for the off-diagonal terms of (\ref{TB_Ham_FT})
differ only by the presence of the factor $\eta_{\bdelta}$:
$$
S'=\sum_{\bdelta} \eta_{\bdelta}e^{i\int_{\br}^{\br+\bdelta} \bw \cdot d\bl}
f_{\br+\bdelta}~~,
$$
and after similar algebra we find
\begin{multline}
S'= e^{i\bk_F^{(1)} \br}
\Bigl[
2\sqrt{2} \delta \sin({\tmpQ}\delta)
\left(
\frac{p_y-p_x}{\sqrt{2}}+\frac{w_y-w_x}{\sqrt{2}}
\right)F\\
-
\delta^2\cos({\tmpQ}\delta)
((p_x+w_x)^2-(p_y+w_y)^2) F+{\ldots} 
\Bigr],
\end{multline}
Using the expressions for $S$ and $S'$, in the leading order  we obtain
\begin{equation}
H_{lin} = v_F \frac{\Pi_x+\Pi_y}{\sqrt{2}}\sigma_3
+v_{\Delta}\frac{\Pi_y-\Pi_x}{\sqrt{2}}\sigma_1
+v_F \frac{v_x+v_y}{\sqrt{2}}
\label{H_TB_lin}
\end{equation}
where the effective velocities $v_F$ and $v_{\Delta}$ are given by the
zero-field expressions given in the next subsection
(\ref{tbvfvd}), $\bv$ is the superfluid velocity, and
$\Pi_i = p_i+a_i$ is the generalized momentum with $\ba=(\bv_A-\bv_B)/2$
describing the vector potential due to an array of Aharonov-Bohm
$\pi$-fluxes located at vortices of subset $A$ and similarly
$-\pi$-fluxes at vortices of subset $B$.
Since the only length in this effective low-energy Hamiltonian 
(\ref{H_TB_lin}) is the magnetic length $l$, it immediately follows 
\begin{equation}
H_{lin}(\br,l, v_F, v_{\Delta}) = \frac{\hbar v_F}{l} H_{lin}(\br/l,
v_F/v_{\Delta}),
\label{H_scaling}
\end{equation}
and, consequently, the spectrum of this Hamiltonian must satisfy the Simon-Lee
scaling relation for the low energy eigenstates:
\begin{equation}
E_{n}(\bk) = \frac{\hbar v_F}{l} E^*_{n}(\bk l, v_F/v_{\Delta})~~.
\label{simonleescaling}
\end{equation}
\subsection{Zero field spectrum}
Let us briefly summarize the properties of the spectrum in the absence of
a magnetic field, which is obtained from (\ref{TB_Ham_FT}) by setting the
field-induced factors $\exp({i\cA_{\br\br'}})$ and
$\exp({i\cV^{A(B)}_{\br\br'}})$ to zero.  The excitation spectrum in this case is
$$
\epsilon_{\bk}=\pm\sqrt{\xi_{\bk}^2+\Delta_{\bk}^2}, \quad {-\pi/\delta\le
k_{x,y}<\pi/\delta}
$$
where
$
\xi_{\bk}= -2t(\cos k_x \delta+\cos k_y \delta)-\mu
$
and the $d$-wave pairing gap function is
$
\Delta_{\bk}=2\tmpDelta(\cos k_x\delta-\cos k_y \delta)
$.
The four nodal points ($i=1,{\ldots},4$) of the spectrum are located at
$\bk_{F}^{(j)} = (\pm {\tmpQ}, \pm {\tmpQ})$, where
\begin{equation}
{\tmpQ}=\frac1{\delta} \arccos\left(-\frac{\mu}{4t}\right).
\label{defQ}
\end{equation}
All four nodes merge at $\mu=\pm 4t$, while at $\mu=0$ the inter-nodal
separation is the largest. 
The  dispersion in the vicinity 
of each node can be approximated as
 $\eps_{\bk} = \sqrt{v_F^2 \delta k_{\perp}^2 +v_{\Delta}^2 \delta
k_{\parallel}^2 }$, where $\delta k_{\perp}$($\delta k_{\parallel}$) is
the displacement of the momentum from a nodal point 
in the direction perpendicular
(parallel) to the Fermi surface, and the  effective velocities are
\begin{align}
v_F& = 2\sqrt{2} a \sqrt{1-\left(\frac{\mu}{4t}\right)^2}\; t\\
v_{\Delta}&=2\sqrt{2}a  \sqrt{1-\left(\frac{\mu}{4t}\right)^2}\tmpDelta~~.
\label{tbvfvd}
\end{align}
\subsection{Spectrum in a magnetic field: $\mu=0$ (known results)}
The structure of the spectrum in a finite magnetic field is
a great deal more complex. In all cases, the 
spectrum at energies much larger than
$\tmpDelta$ evolves to a set of 
sharp -- conventional Schr\"odinger, as opposed to
Dirac -- Landau levels, while at low energies is characterized by 
strongly dispersive energy bands. For  concreteness, we will
consider a square lattice of vortices oriented as shown in  Fig.
\ref{unitcell} with a unit cell of minimal area, which  contains two
vortices, labelled $A$ and $B$. In addition, we will only study
the values of magnetic field that correspond to the symmetric placement
of vortices within plaquettes  -- this requirement restricts the values of
magnetic length $l$ to even integers in the units of lattice spacing. 

Let us first recall the results for a fully particle-hole symmetric system
\cite{vmft, vmt, vm_index}, corresponding to $\mu=0$. Due to a special symmetry
($\psi_{\br}\to (-1)^{(x+y)/\delta}\psi_{\br}$) of this case, 
the spectrum is doubly degenerate for all momenta $\bk$.
As shown in  Ref. \onlinecite{vm_index}, when the 
center of inversion for the  vortex 
lattice coincides with a site of the atomic lattice -- a
situation realized for magnetic lengths  $l/\delta\equiv 2 \pmod 4$-- the 
spectrum contains  {\em eight}  Dirac nodes ({\em sixteen} zero energy states):
two degenerate nodes at each of the four  momenta
$\bk=(\pm \frac{\pi}{2l}, \pm \frac{\pi}{2l})$.
This is quite unusual non-perturbative result, since it
suggests that for arbitrarily small fields giving 
rise to a vortex lattice, the number of zero modes
is doubled compared to the four nodes of the zero-field problem,
provided the magnetic length has the correct commensuration with the
tight-binding lattice spacing.

In the opposite case $l/\delta\equiv 0 \pmod 4$, the symmetry does not
demand the zero modes, and, as was found numerically in
Ref. \onlinecite{vmft}, the field-induced gap in 
this case scales as $l^{-1}$ as a function
of magnetic length within the regime $H_{c1}\ll H\ll H_{c2}$. This
 result is also surprising, since for the  lowest energies,
one expects to recover the Simon-Lee scaling of 
the linearized problem (\ref{simonleescaling}).

The existence of the nodes for commensuration $l/\delta\equiv 2
\pmod 4$ might suggest that the spectrum of the linearized problem given by
$E^*$ in  (\ref{simonleescaling}) is gapless -- this is also a result 
obtained earlier \cite{ft_BLOCH,mhs} from the analysis of the
linearized Hamiltonian. This conclusion  in turn would have required 
that in the expansion of the overall field-induced gap
$$
\Delta_m = \alpha_1 \frac{1}{l}+\alpha_1 \frac{1}{l^2}+{\ldots}
$$
the leading $1/l$ term is absent, and only small gaps of order $1/l^2$
should generally appear from the terms that were left 
out in the process of linearization. 

The large gaps whose size scales as $1/l$ for
weak magnetic fields in a particle-hole 
symmetric situations at commensuration factors $l/\delta\equiv
0\pmod{4}$, as was found in Refs. \onlinecite{vmft,vmt}, therefore
come entirely unexpected. These large gaps were interpreted as 
as the effect of the inter-nodal
interference, and such effects for the deliberately distorted lattice were
indeed found to be suppressed. It was argued that in realistic situations 
 a weak disorder in vortex positions or the one due to 
impurities  will suppresses these
interference effects. Yet, the
following  questions remain to be answered: 
First, is this situation specific only to  a particle-hole symmetric
system ($\mu=0$)?
Second, how could the scaling relation (\ref{simonleescaling}) 
be explicitly violated, 
even for an ideal periodic lattice?  Finally, how -- if at all -- the 
Simon-Lee scaling can be recovered in the tight-binding 
problem without introducing the disorder
explicitly?

\subsection{The spectrum for general $\mu$}
We start by  answering the first of these question and consider 
the spectrum for
non-particle-hole symmetric systems ($\mu\neq 0$). 
We show below that the rather
involved  behavior displayed by the spectrum as a function of $\mu$ and $l$
in fact follows a simple universal pattern when expressed
in terms of suitably rescaled
variables. For the square lattice of vortices considered here,
the analysis is further simplified due to the 
sixteen-fold  symmetry of  the dispersion $E_{n\bk}$ within the 
Brillouin zone illustrated in Fig. \ref{unitcell}. Furthermore, using a transformation $\psi_{\br} \to
(-1)^{(x+y)/\delta} \psi_{\br}$ and the symmetry of the spectrum at each
$\bk$ as a function of energy (see Appendix), it is easy to show that the spectra
at values of chemical potential $+\mu$ and $-\mu$ are identical, with or
without magnetic field, and 
therefore in what follows  we assume $\mu>0$.

\begin{figure}[tbh]
\centering
\includegraphics[width=\columnwidth]{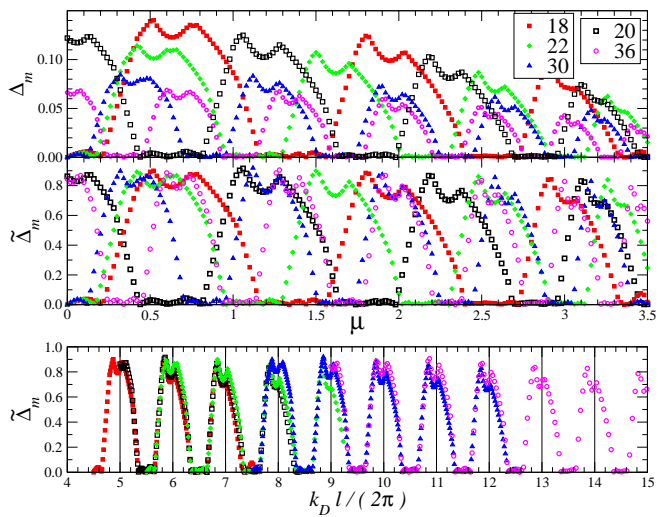}
\caption{\label{fig:TBoscgap} Upper panel: The overall 
gap $\Delta_m$ induced by vortex
lattice as a function of chemical potential $\mu$. The parameters are
$\tmpDelta=t$, $l/\delta=18,22,30$ (solid symbols) and $l=36\delta$ (open symbols). 
Center panel: the gaps are renormalized by $l/v_F$ according 
to the Simon-Lee prescription: $\tilde{\Delta}_m= \Delta_m l / \hbar v_F$.
In the limit of small magnetic field the result should have been independent of
$\mu$ in direct contradiction with explicit numerical evaluation shown here. 
Lower panel: instead of the chemical potential $\mu$, the horizontal axis
represents $\tmpQ l/(2\pi)$. In the limit of low magnetic fields ($l\gg \delta$) all curves
representing dependence of $\tilde{\Delta}_m$ on $l  \tmpQ $ collapse onto a
$2\pi$-periodic function. For fixed $l$, deviations from this universal scaling
are the largest for $\mu$ close to the bottom of the tight-binding band, where
the Fermi surface is small and the validity  condition for linearization
($l k_F \gg 2\pi$) is violated.}
\end{figure}
%
\begin{figure}[tbh]
\centering
\includegraphics[width=\columnwidth]{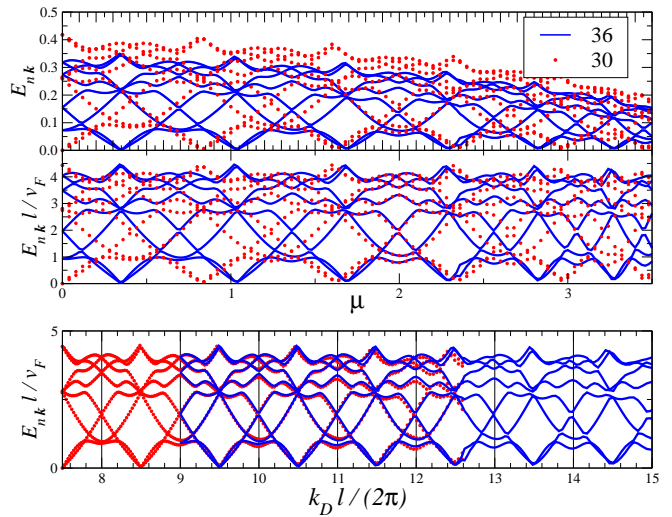}
\caption{\label{fig:TBosc} Upper panel: The dependence of the lowest eight
positive energy bands $E_{n\bk}$  at fixed 
momentum $\bk_0 l =(\pi/2,\pi/2)$ on chemical potential
$\mu$.  The parameters are $\tmpDelta=t$, $l/\delta=30, 36$. Note that the two
values of magnetic length $l$ correspond to the gapped ($l/\delta=4n$) and gapless
($l/\delta=4n+2$) families at half-filling ($\mu=0$). Center panel: the energy
levels at $\bk_0$ are rescaled by $l/v_F$.
Lower panel:  Instead of the chemical potential $\mu$, the horizontal axis
represents $l  \tmpQ/(2\pi)$. In the limit of low magnetic
fields ($l\gg \delta$) all curves
representing dependence of $\tilde{\Delta}_m$ on $\tmpQ l$ collapse onto a
$2\pi$-periodic function.}
\end{figure}
The field-induced gap for $v_F=v_{\Delta}$ is plotted as a function
of the chemical potential  $\mu$ in the upper 
panel of Fig. \ref{fig:TBoscgap} for several
values of magnetic length $l$. At all magnetic fields the 
dependence on $\mu$ displays a characteristic
oscillatory behavior. For the family of magnetic lengths 
$l=(4n+2)\delta$,  the spectrum is gapless at
$\mu=0$, in accordance with the previous results\cite{vmft},
and for a finite fraction of one cycle of oscillations in $\mu$ the
field-induced gaps are extremely small, their size 
quite possibly set by $l^{-2}$ in the scaling limit: we were 
not able to definitely establish the
scaling behavior due to the smallness of the gaps in this regime. 
Then the gaps increase and remain large -- of order $\hbar v_F/l$ 
--  for about a third of the cycle, until eventually again
turning to zero. This cycle is then repeated over 
and over.  For $l=4n\delta$ the
only difference is that the cycle is offset by
half a period in $\mu$. The overall slow
decrease of the average gap size for large values of $\mu$ follows directly
from the Simon-Lee scaling (\ref{simonleescaling}) as  $v_F$ decreases with
$\mu$ (see Eq. (\ref{tbvfvd})). To account for the expected 
Simon-Lee scaling, the central panel of Fig. \ref{fig:TBoscgap}
shows the rescaled gap 
$$
\tilde{\Delta}_m = \frac{\Delta_m l }{\hbar v_F}
$$
as a function of $\mu$. If Simon-Lee scaling in 
its original form were exact, one would expect to
see no dependence $\mu$. Instead, for any given field value, the rescaled gap
$\tilde{\Delta}_m$ itself exhibits oscillatory behavior. 

Still, comparing the  upper
panel of Fig. \ref{fig:TBoscgap} one must conclude
 that $\tilde{\Delta}_m$  is a step forward compared to  
 $\Delta_m$; {\em on average} the curves
representing different magnetic field and different values of $\mu$ look
almost identical. Although the detailed analytic theory 
of the ``interference effects'' remains a challenge for the
future, the essence of such interference is vividly 
illustrated by replotting the family of  $\tilde{\Delta}_m$ as function of
$l  \tmpQ/(2\pi)$.  So rescaled,
all curves collapse into a single universal periodic
function shown in the third panel of Fig. \ref{fig:TBoscgap}. 

We find that the above oscillatory behavior is not specific to the
field-induced gap function; the dependence of 
the entire spectrum $E_n(l\bk)$ is
characterized by similar behavior. As an example, Fig. \ref{fig:TBosc}
displays the eight lowest
energy levels at $\bk=(0,0)$  for $l=30\delta$ and
$l=36\delta$, representing
two families of magnetic fields. The central panel shows the 
 energy levels rescaled by $v_F/l$, while the bottom panel shows that the
remaining oscillations of the rescaled spectrum fall onto a universal
periodic curve if plotted as functions of $\tmpQ l$ rather than $\mu$.

\begin{figure}[tbh]
\centering
\includegraphics[width=\columnwidth]{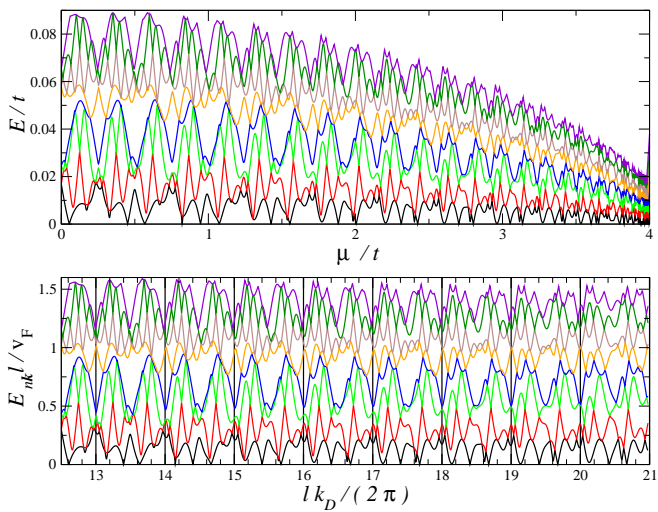}
\caption{\label{fig:TBosc_0.2} Upper panel: The dependence of the lowest eight
positive energy bands $E_{n\bk}$  at fixed 
momentum $\bk_0 l =(0,0)$ on chemical potential
$\mu$.  The parameters are $\tmpDelta=0.2t$, $l=50\delta$.
Lower panel:  Instead of the chemical potential $\mu$, the horizontal axis
represents $l  \tmpQ/(2\pi)$. Only the range $0<\mu<3.5t$ is
plotted in the lower panel.}
\end{figure}

The pattern just described  is not restricted to the isotropic case
$\alpha_{\Delta}=v_F/v_{\Delta}=1$ and 
holds for all $\alpha_{\Delta}$ we 
checked ($2$, $5$, $10$, $20$, $50$).
For large anisotropies, the deviations from the scaling behavior become
more pronounced at smaller values of $l$ due to violation of the $\hbar
\omega_c \ll \tmpDelta$ condition, which translates
to  $l/\delta\gg \sqrt{\alpha_{\Delta}}$ -- this condition is
necessary if we are to treat magnetic field as a ``small''
perturbation of a zero-field $d$-wave superconductor.
Fig. \ref{fig:TBosc_0.2} illustrates the spectrum at $k=0$ for
the anisotropic case $\alpha_{\Delta}=5$.
In all cases, for sufficiently weak magnetic fields, the spectrum still 
exhibits  $2\pi$-periodic oscillations in $\tmpQ l$ of amplitude
 $C \hbar v_F/l$, where the numerical prefactor $C$ is only a function of
$t/\tmpDelta$. 
We stress that the oscillatory part of the spectrum is of
the same order of magnitude as its smooth, ``envelope'' dependence,
and is not smaller than this average Simon-Lee envelope part in any
sense. These results can be summarized in the following scaling form:
\begin{equation}
E_{n\bk} = \displaystyle\frac{\hbar v_F}{l} \cE_n(\bk l, t/\Delta,
\tmpQ l),
\label{modsimonlee}
\end{equation}
where $\cE$ is a universal 
dimensionless function, which is $2\pi$-periodic with respect to its last
argument. This scaling, which combines the oscillations with respect to both
magnetic field and the chemical potential,  holds for all values of chemical potential, except
when $\mu$ is very close to the bottom of the band -- in this case the
scaling was studied by Vafek and  Te\v{s}anovi\'c\cite{vtscaling}.

The origin of this oscillatory behavior can be 
traced back to the linearization procedure.
Recall that the matrix elements of the FT-transformed, but yet
non-linearized Hamiltonian  ${\cal H}$ (\ref{TB_Ham_FT})
are evaluated in the plane wave basis. By inspection, in the limit 
$l  \tmpQ \gg 1$, the leading term of this infinite  matrix
${\cal H}_{\bk\bk'}$ appears to have a block-diagonal
form, with each of the four blocks corresponding to separate nodes.
Within each block, only the leading order approximation in $l^{-1}$
is kept, while the block-offdiagonal (inter-nodal) matrix
elements as well as the higher-order corrections to the 
block-diagonal (intra-nodal) matrix elements, which 
superficially scale at worst as $l^{-2}$ , are neglected.
Therefore, by construction, the matrix elements of each block
precisely coincide with the matrix
elements of the linearized Hamiltonian. What 
we found, however\cite{vmt,mt_SAE}, is that the situation is not this simple:
this description is necessarily 
incomplete due to the singular character of the linearized problem 
wavefunctions near vortices. Consequently, the linearized Hamiltonian
requires self-adjoint extensions, obtained
by imposing additional boundary conditions
near vortices, which themselves are ultimately determined 
by those ``higher-order terms'' which were dropped in the oversimplified
analysis.

A straightforward 
way to appreciate the significance of these subdominant terms is
actually to recompute the matrix elements of the full Hamiltonian -- not
in the plane-wave basis as one naturally would within a pedestrian
perturbation theory -- but with
respect to the exact eigenfunctions of the 
linearized problem\cite{thanksov} with fixed boundary
conditions at vortex locations $\{\theta\}$. The
wavefunctions of the linearized Hamiltonian 
diverge near vortices\cite{mt_SAE} as
$r^{-1/2}$, and therefore the eigenfunctions of the linearized
Hamiltonian at each node can be written as
\begin{multline}
\Psi_{\bk, n}^{(j)}(\br) = e^{i \bk_F^{(j)}\cdot \br}
\left(
\frac1{\sqrt{l|\br-\bR_A|}}
\chi^{(jA)}_{\bk, n}(\br/l)\right.\\ \left.
+\frac1{\sqrt{l|\br-\bR_B|}}
\chi^{(jB)}_{\bk, n}(\br/l) +\frac1l f_{\bk, n}(\br/l)\right),
\label{wf_div_reg}
\end{multline}
where $i$ labels the nodes $1,\overline{1}, 2$, and $\overline{2}$, $\bk$ is
the Bloch momentum,  $n$ is the band index, 
and dimensionless continuous functions
$\chi$ and  $f$ are such that the expression in the brackets is
Bloch-periodic. Note that if the wavefunctions
contained only the regular part, the terms retained in the linearized
Hamiltonian would be of the order $\hbar v_F/l$, while the non-linear terms
such as $m \bv_{A(B)}^2/2$, $\ba^2$ etc,  would contain {\em an additional}
factor of $(k_F l)^{-1} \ln
(l/\xi)$, and could have been safely omitted. The presence of
the divergent part of the wavefunctions described by the first part of
(\ref{wf_div_reg}), however, changes the situation. Let us evaluate again the
structure of the matrix elements $\langle\Psi^{(j')}_{n\bk}|{\cal
H}|\Psi^{(j)}_{n'\bk'}\rangle$ between the states at momenta $\bk$ and $\bk'$ differing by
reciprocal lattice vector $\bG$. The {\rm intra-nodal} matrix elements with
$j=j'$  to the leading order  are just the eigenstates of the
linearized Hamiltonian $\hbar v_F E_n^*(\bk l) \delta_{nn'}/l$. The
corrections due to non-linear terms, however, quite peculiarly also  exhibit
the same scaling as a function of magnetic length $l$ as we will argue
now. Consider a typical non-linear term $m\bv^2/2$, since $(m\bv)^2$ increases 
near vortices as $1/r^2$ down to distances of
the core size $\sim\xi$, to the leading order the matrix element 
\begin{multline}
\left\langle\Psi^{(j)}_{n\bk}\Bigr| \frac{m\bv^2}{2}\Bigr|\Psi^{(j)}_{n'\bk'}\right\rangle\propto
\int_{\xi}^l  (rl)^{-1/2} \frac{r^{-2}}{m}(rl)^{-1/2} (rdr) \\
\propto
\frac{v_F}{l} \frac1{k_F\xi}
\label{curvatureINTRA}
\end{multline}
has the $1/l$ scaling. Higher-order curvature terms containing higher order
derivative operators and potentials $\ba$ or $\bv$ can be estimated
similarly; corrections to the matrix elements due each successive term
$|a_{i_1}|^{j_1} \partial_{i_2}^{j-j_1}$ in general 
are of the order of $v_F (k_F \xi)^{-(j-1)}/l$. Therefore, the linearization procedure
in the presence of magnetic field is justified when
the condition $(k_F\xi)\gg 1$ is satisfied, and the role of the small
parameter is played by both $(k_F \xi)^{-1}$ and 
$(k_F l)^{-1}\ll 1$, and not only the latter, as is commonly assumed.

Moreover, the ``interference'' terms relating
different nodes ($j\neq j'$) have a similar form and scale as $l^{-1}$:
\begin{multline}
\left\langle\Psi^{(j)}_{n\bk}\Bigr| \frac{m\bv^2}{2}\Bigr|\Psi^{(j')}_{n'\bk'}\right\rangle 
\propto
\frac{v_F}{l} \frac1{k_F\xi}\\
\times\left(C_1e^{i\bR_A\cdot\bG}+C_2e^{i\bR_B\cdot\bG}\right),
\label{curvatureINTER}
\end{multline}
where $\bG=(\bk^{(j)}_{F}+\bk)-(\bk^{(j')}_{F}+\bk')$ and 
coefficients $C_{1,2}$, determined by  the wavefunctions
$\chi^{(j\alpha)}_{\bk,n}$ and $\chi^{(j'\alpha)}_{\bk',n'}$ ($\alpha=A,B$),
depend on $(n,\bk;n',\bk')$ but not on $\bk^{(j)}_F$ or $\bk^{(j')}_F$.
One therefore generally 
anticipates that -- with $k_F \xi$  kept fixed, as in our
model, and other parameters (such as  $l$ or $\mu$) freely varied --
the spectrum will undergo a complicated evolution,
even at the leading order in $l^{-1}$ (c.f. Ref. \onlinecite{vmft}), due to
the inter-nodal contribution enhanced by the singular character of
wavefunctions near vortices. 
Note that in the tight-binding lattice model with the nearest
neighbor hopping terms only, we are precisely in this situation: in our
simplified model, where no self-consistency condition is employed, $k_F\xi$
is a fixed number of order $1$ since 
the role of the cut-off $\xi$ is played by the 
lattice spacing $\delta$, and
$k_F$ is bounded by $(\pi/2)\delta ^{-1}$. 
Even when the self-consistency condition is employed, as long
as the uniform system
is described by nearest neighbors only, the 
ratio $v_F/v_{\Delta}$ automatically
fixes $t/\tmpDelta$; for a fixed anisotropy of $d$-wave nodes,
$k_F\xi$ is bounded by a number of the order of $\alpha_{\Delta}$
since $v_F/v_{\Delta}\sim  t/\tmpDelta\sim k_F \xi$ and $\xi\sim \delta$, 
and therefore
the simple Simon-Lee scaling limit will be difficult to reach in the
strict sense. Of course, one may introduce the next-nearest neighbors 
and further hopping terms in order to 
optimize parameters and maximize $k_F\xi$ while retaining the fixed value of
$\alpha_{\Delta}$. In this case, the amplitude of the oscillations will
still scale as $l^{-1}$, albeit with a suitably reduced prefactor.

The above  ``interference'' pattern of the spectrum 
for a moderately large  $(k_F\xi)$ is expected to
have a periodic structure, depending on the commensuration of the nodal
wavevectors and a  magnetic length. Consider a change in chemical potential
$\mu$ or other parameters that result in a displacement of 
nodal points at the Fermi surface. If the difference 
$(\bk^{(j)}_{F}-\bk^{(j')}_{F})\cdot (\bR_{B}-\bR_{A})$ changes by a
multiple of $2\pi$, then the amplitudes of the  matrix elements in
(\ref{curvatureINTER})
between $(n,\bk)$ and $(n',\bk')$, which 
determine the leading order perturbative
corrections to the energy spectrum, are the same to the leading order in
$l^{-1}$, apart from the  prefactor $(k_F \xi)^{-1}$. Thus, in addition
to overall the Simon-Lee ``$v_F/l$'' scaling,  the
spectrum has periodic oscillations determined by the commensuration of the
inter-nodal momentum $\bk^{(j)}_F-\bk^{(j')}_F$ and the difference 
$\bR_A-\bR_B$. More precisely, the spectra for two sets of parameters 
will be  similar whenever the nodal 
points $(\pm {\tmpQ}, \pm {\tmpQ})$ and $(\pm {\tmpQ}',
\pm {\tmpQ}')$ satisfy the condition
\begin{equation}
{\tmpQ}l -{\tmpQ}' l' = 2\pi n,
\label{theoretical_periodicity}
\end{equation}
where $n$ is an integer.
This is equivalent to Eq. (\ref{modsimonlee}) 
surmised from the numerical solution.

\begin{figure}[tbh]
\centering
\includegraphics[width=\columnwidth]{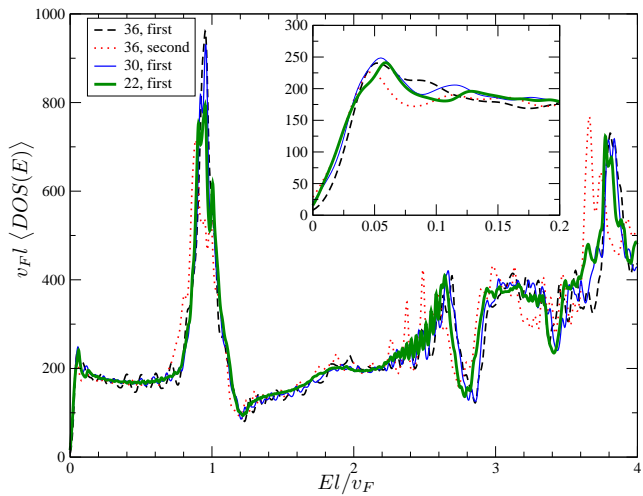}
\caption{\label{fig:TBaveDOS} Full DOS for $\alpha_{\Delta}=1$ averaged over
one cycle of oscillation in $\mu$ for several values of magnetic field
characterized by magnetic length $l$. The density of
states was computed for all values of $\mu$ on a mesh of size
 $\delta\mu=0.02t$ for $l=22\delta, 30\delta$ and
 $\delta\mu= 0.01t$ for $l=36\delta$.
 The averaged DOS is shown
for the first period of oscillations $\mu\in(0,4t\sin(2\pi \delta/l))$ and for the
second period $\mu\in(4t\sin(2\pi \delta/l),4t\sin(4\pi \delta/l))$. The inset shows
the enlarged low-energy part of the figure. }
\end{figure}

A remarkable feature of the oscillations seen in Figs. \ref{fig:TBoscgap} and
\ref{fig:TBosc} is that the frequency of oscillations in $\mu$ grows rapidly 
with magnetic field.
Incidentally, this suggests a way of incorporating 
the effect of weak disorder, which is 
expected to suppress the oscillations, in a relatively
simple manner: weak disorder in a full calculation is 
equivalent to weakly modulated $\mu(\br)$.
On the other hand, since the spectrum is a rapidly oscillating function of
$\mu$, only the quantities averaged over one period of
oscillations are of interest.  For a typical value of magnetic 
field ($\sim 1$ Tesla) the period of oscillations can be 
estimated to be of order $10$~meV.
Thus, if a random impurity potential $\mu(\br)$ is 
of comparable magnitude or larger,
only the averages of measurable physical quantities over 
a period of oscillation are observable.
In  Fig. \ref{fig:TBaveDOS} the density of states (DOS) averaged over the first
and the second periods close to half-filling are shown for a variety of
magnetic fields. At low energies, DOS is linear in energy and
should be associated with the nodal  
structure of the spectrum {\em on average}.

%

We end this section by alerting the reader to the fact that,
although we have  performed detailed numerical calculations for the
nearest-neighbor hopping  model only, where the strict reduction to the
simple linearized description is hampered by the 
enhanced effects of the inter-nodal
interference and curvature terms, we expect that in a 
more elaborate (and more realistic) model, with longer range hoppings,
where the condition $k_F\xi \gg 1$ is better satisfied,
the linearized effective theory does indeed provide a 
quantitatively faithful description
of the low energy sector of the theory. 
In that case, the singular nature of
the potential due to vortices still requires special care,
but such care can be administered  
by constructing suitable self-adjoint
extensions of the linearized  Hamiltonian, as belabored in Ref.
\onlinecite{mt_SAE}, where a detailed discussion of the
Dirac-Bogoliubov-deGennes quasiparticles in singular vortex potentials is
presented.

\begin{figure}[tbh]
\centering
\includegraphics[width=\columnwidth]{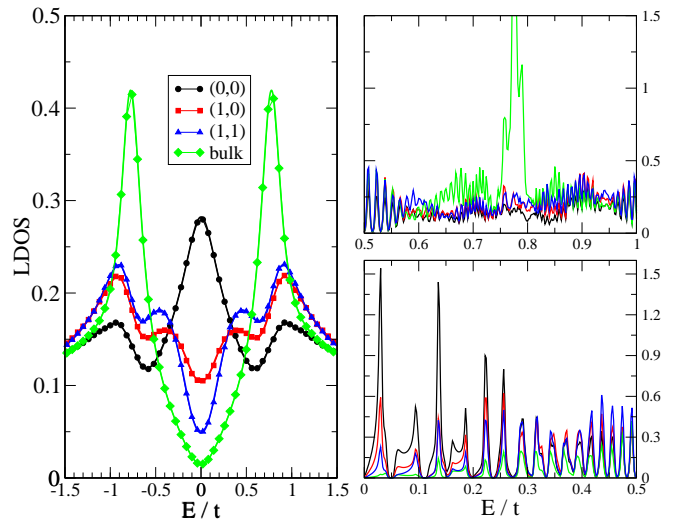}
\caption{\label{fig:largescaleldos} Left panel: TLDOS at $\mu=0$  in a
mixed state at 4 representative points for $l=50\delta$ (lines)  and
$l=38\delta$
(symbols). 
Far from the vortices, the TLDOS of a uniform $d$-wave
superconductor is recovered. At four corners of the plaquette containing the
vortex the TLDOS exhibits the zero-bias peak \cite{wang_macdonald}, while at
the next-nearest and the next-next-nearest sites, the TLDOS develops peaks
at sub-gap energies. Not only the position of these peaks, but also the
thermally broadened TLDOS at all energies does not depend on magnetic field
(length) provided that the temperature is larger than a typical width of a
band ($T=0.05t$ here). This large-scale, ``high-energy'' behavior of the
thermally broadened TLDOS should be contrasted with the stark dependence of
the low-energy features on $l$, commensuration effects etc, which we focused
on in the previous section. This ``fine'' structure, which corresponds to
true TLDOS is shown on the right, where TLDOS is plotted for
$l=22\delta$. As we described earlier, the latter is generically  gapped with the
gap scaling as $l^{-1}$ or, as in the example shown in the right panel,
is linear  for special commensuration between the
magnetic length and  the Fermi momentum. 
}
\end{figure}

\section{TLDOS near vortices and the missing zero-bias coherence peak}
\label{section:ZBCP}
So far we were describing the details of the spectrum at the lowest
energy scale set by $\hbar v_{F}/l$. Now we turn to large-scale properties 
of the TLDOS $g(\br,E)$
-- a quantity of direct
interest in the STM experiments, which can be expressed through the
eigenstates of the BdG Hamiltonian $(u_{\br}, v_{\br})$ as
\begin{multline}
g(\br, E) \propto \sum_{n\bk} \Bigl(|u_{n\bk}(\br)|^2 f'(E-E_{n\bk})\\
+|v_{n\bk}(\br)|^2 f'(E+E_{n\bk})\Bigr),
\label{tldos}
\end{multline}
where $\br$ denotes the site of the TB lattice and $E$ is the bias. While the results at these large
energies are less universal  and depend to a much larger degree on the
band structure, the spatial profile of the order
parameter etc., certain  qualitative features turn out to be rather
robust and will be discussed in this section.

A typical dependence of the TLDOS on bias for 
particle-hole symmetric case $\mu=0$ is shown in
Fig. \ref{fig:largescaleldos} for a set of representative points of
the tight-binding lattice.
First note that the thermally broadened TLDOS is essentially  field-independent
once the temperature exceeds the typical width of the
field-induced bands, which varies from $C(\alpha_{\Delta})\hbar v_{F}/l$ at low energies,
 where the spectrum is strongly dispersive
to $\hbar \omega_c \sim 2\pi t/l^2$ at energies
larger than $\Delta_c$ (see Refs. \onlinecite{ft_BLOCH, mhs,vmft}). As an example, compare the
TLDOS for $l=50\delta$ (lines) and $l=30\delta$ (symbols), which are shown in the left panel of Fig.
\ref{fig:largescaleldos} for temperature $T=0.05t$.

Far from the vortices the TLDOS is similar to the zero magnetic field
result, and quite naturally the deviations grow 
progressively stronger as one approaches a
vortex. As found by Wang and MacDonald\cite{wang_macdonald}, 
on four cites surrounding the plaquette
with the vortex the thermally broadened TLDOS has a pronounced 
maximum at $E=0$ as a function of the 
applied bias voltage, which is called the zero
bias conductance peak (ZBCP). Note, however, that ZBCP
 appears only after thermally broadening the TLDOS, 
which in its original form is either gapped
at general $\mu$ or has a linear dependence on the energy
for a discrete set of  $\mu$  such as $\mu=0$ 
and $l/\delta=2\pmod{4}$ as discussed in the previous
section (see the right panel of Fig. \ref{fig:largescaleldos}).
To resolve this low-energy gapped or linear 
dependence, however, the temperature must
be smaller than a typical width and separation 
between the bands. We stress that the ZBCP does not
correspond to any ``localized state'': many narrow
Bloch bands (see right panel of Fig. \ref{fig:largescaleldos}) merge 
into a peak after thermal broadening.

Importantly, the situation is very different 
on the next nearest and next-next nearest 
neighbor sites around a vortex: there
the local density of states exhibits peaks at
energies $\approx \pm \Delta_c/2$, where $\Delta_c$ denotes the coherence
peak energy in a uniform system. Note that these peaks share 
several similarities with the  sub-gap features observed in
experiments, namely, the energy of 
these sub-gap peaks is independent of magnetic
field (see left panel in Fig.  \ref{fig:largescaleldos}) and 
it also increases
with $\Delta_c$. Again, these peaks do not correspond to any
``localized state(s)'', as many narrow bands contribute to the peaks after
thermal broadening of the LDOS.
Although in itself this observation does not quite suffice to explain
the absence of ZBCP in experiments, it does suggest that the
experimentally observed TLDOS might be reproduced quantitatively by
considering a standard $d$-wave vortex on a lattice with some relatively minor
modification, rather than invoking the appearance of 
additional order  parameter(s) within vortex core(s).

A hint of such a minor modification, which could suppresses the ZBCP 
at the four sites closest to the vortex, comes from a recent 
analysis of the  dopant oxygen atoms in BSCCO.
As was noticed by Nunner {\em et al.} \cite{nunner}, 
the nature of spatial correlations
between the position of the oxygen atoms \cite{mcelroy} and features 
observed in the
TLDOS, indicates that the strength of 
the electron-electron effective pairing
coupling constant, and therefore also the  
magnitude of the $d$-wave superconducting 
gap function in BSCCO, are both strongly susceptible to local variations.
Variations of  $\Delta$ by a factor of two or more on a scale of a few lattice
spacings, which are basically never seen in conventional superconductors, are
routinely observed in BSCCO and other cuprates. In a zero-field case it was
found\cite{nunner} that these
modulations of the $\Delta$ are likely to be 
caused by dopant atoms; while the detailed
microscopic origin is not clear at this point, it is conceivable that
dopant atoms cause local distortions of the atomic lattice and cause
spatial variation of the superexchange interaction or other interaction
important for superconductivity.
%
\begin{figure}[tbh]
\centering
\includegraphics[width=\columnwidth]{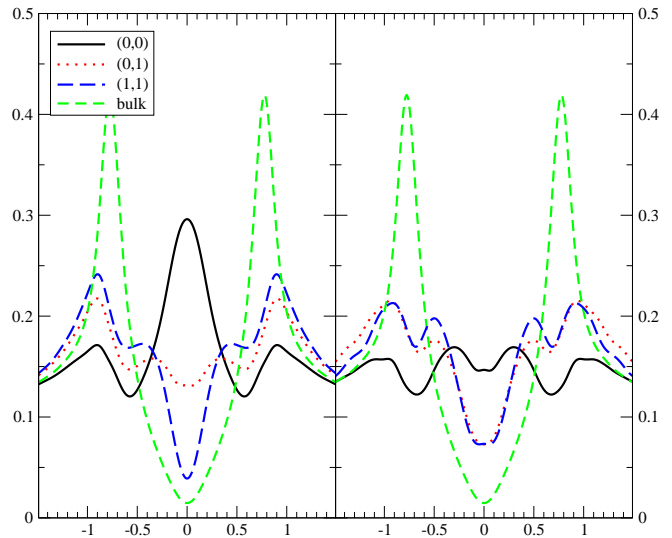}
\caption{\label{fig:ldosIMP} Left: the thermally broadened
TLDOS for a $d$-wave superconductor with
spatially constant amplitude of the gap function, except on four bonds surrounding
each vortex, where the gap function
 is set to zero. The parameters are: $\tmpDelta=0.2t$, $l=38\delta$,
$\mu=0$. The temperature is chosen as $T=0.05t$, and 
the origin  $(0,0)$ denotes the upper-right corner 
of a plaquette containing the vortex. Note that the nearest sites to
the vortex exhibit the ZBCP, while sites $(1,0)$ and $(1,1)$ have
sub-gap(sg) peaks at non-zero energies $~0.5\Delta_{sg}$. Compared to the case where there
is no suppression of $|\Delta_{\br\br'}|$ near vortex core, the ZBCP is
slightly enhanced.
Right: Same, but bond variables on four bonds around each
vortex are increased by a factor of $3$. Note that
this change affects most dramatically the nearest sites $(0,0)$ to
the vortex, where the ZBCP is suppressed and two maxima at sub-gap
energies develop.}
\end{figure}
%
Since vortices are expected to  be pinned by impurities, the 
profile of the order parameter near a vortex
would consequently differ from that in an
ideal model considered so far. Furthermore, even if the
vortex is not pinned by an impurity, it may distort the lattice and cause
variations of the order parameter near its core that are not described in
the canonical BdG scheme, where the pairing interaction constant is assumed
to be spatially uniform. 

While at this stage the above reasoning in the context of cuprates
is only a speculation, it is still a useful phenomenology to
examine more closely the effect
of such modulation of the $d$-wave gap
function near the vortex core. The main results
are summarized in Fig. \ref{fig:ldosIMP}. First, consider a  suppression of 
the gap function all the way to zero on  four 
bonds surrounding plaquette with the vortex;
the result is shown in the left panel of Fig. \ref{fig:ldosIMP}. 
The 
ZBCP is strongly enhanced, while other features are modified
only a little. In the opposite case 
(see right panel of Fig. \ref{fig:ldosIMP}),
when the  magnitude of $d$-wave gap function
is locally enhanced, the zero-bias conductance
peak is  strongly suppressed, the spectral weight is transferred to the
$\Delta_c/2$ sub-gap states, and TLDOS acquires a form 
rather similar to that observed
in experiments. The suppression of the ZBCP is even stronger if the bonds
with enhanced gap function extend further. Note that in experiments
the sub-gap peaks have an energy lower than $\Delta_c/2$: in BSCCO Pan et al.
reported\cite{pan} the sub-gap(sg)
features at $\Delta_{sg}=7$ meV for the samples
with the coherence peak at $\Delta_c = 32$ meV, 
while Hoogenboom {\em et al.}
\cite{hoogenboom_physica} cite 
values $\Delta_{sg}=\pm 14$ meV and $\Delta_c =\pm
45$ meV. In the earliest samples of 
YBCO, where these low-bias features were observed,
Maggio-Aprile {\em et al.} reported
$\Delta_{sg}=\pm5.5$ meV for system described
by $\Delta_{c}= \pm 20-25$ meV. In all cases, the ratio $\Delta_{sg}/\Delta_c$
ranges from $0.20$ to $0.33$. Since we used the
simplest tight-binding model described by only two parameters $(t,\mu)$, the
numerical discrepancy between the result $\Delta_{sg}/\Delta_c\approx 0.5$
and the experimentally observed ratios is not unexpected.

Finally, we comment on the $4\times 4$ modulations observed in the vicinity
of vortex cores \cite{hoffman, fischer2005}. Such modulations are likely to
be caused by strong fluctuations of the $d$-wave order parameter, which are
believed to become enhanced near vortices. Explanation of such effects is
clearly beyond the scope of the present paper based on a simple mean-field
formulation. It has, however, been argued by several groups
\cite{handong,t4x4, mt4x4,sachdev} that even in the 
absence of magnetic field the enhanced phase fluctuations
of the $d$-wave gap function result in a 
broken translational symmetry with modulated local
average of the gap function. This provides an alternative mechanism
of the modulations in the absolute value of $\Delta$.

\section{Single vortex problem}
\label{section:single}
In this section, we study a problem
of nodal BdG quasiparticles in presence
of a {\em single vortex} piercing a superconducting grain or droplet of size
$L\times L$, where $L$ relates to an external
magnetic field in such a way that exactly one
superconducting flux $\Phi_0=hc/(2e)$ fits into the system. A continuum
version of a similar problem
was considered in Ref. \onlinecite{ft_DPLUSID},
where the delocalized character
of the core quasiparticle states was established. 

The present problem is technically somewhat easier to handle
than the periodic {\em array} of vortices of the previous sections; however,
there are certain general features common to both situations.
In particular, the anomalous enhancement of 
the inter-nodal interference and curvature terms
by the $r^{-1/2}$ increase of the wavefunctions near 
vortex location within the linearized framework still influence the spectrum,
albeit now in a less dramatic 
fashion -- unlike the translationally-invariant case
considered in previous sections, 
quasiparticles within an
isolated superconducting grain have energy levels that to the
leading order in $L^{-1}$  exhibit oscillations as function of $\tmpQ L$ 
even in the absence of magnetic field. 
When a magnetic field is turned on,  the singularities 
of the wavefunctions near the vortex result in the halving of the
oscillation period.

The starting point for description of the quasiparticles inside such a 
superconducting grain is
still the Hamiltonian (\ref{TB_Ham_start}), except now the BdG
wavefunctions $(u(\br),v(\br))$ are required to vanish outside the grain.
Alternatively, all bond variables such as $\Delta_{\br\br'}$ 
or $t_{\br\br'}$ can be set to zero on links along the perimeter of the grain.
The remaining bond variables $\Delta_{\br\br'}$  in principle 
should be determined from the self-consistency conditions, however, just as
in the case of the vortex lattice problem, this approach has a drawback of
depending on the precise form of the microscopic theory and
furthermore on  the precise nature of the boundaries.
Following our justification from the previous section, and 
in order to focus on the
simplest model with the least number of parameters, we
describe the results for the order parameter with a constant amplitude 
$|\Delta_{\br\br'}|$, its phase simply given by the
polar angle around the origin. We verified explicitly that 
after implementing the self-consistent solution of 
the problem using the condition
(\ref{TB_selfconsistency}) we find only small quantitative deviations from
the results described in this section, and no change in the qualitative
conclusions. Although the ``constant amplitude'', ``polar angle'' approximation
for the order parameter is violated near the boundaries of the grain,
its effect on the physics near the vortex appears insignificant.

Before presenting the numerical results, let us start with several 
simple observations.
Consider again the low-energy effective description of $H_{BdG}$.
The singular gauge transformation and subsequent linearization
proceed just as in the case of the vortex lattice
problem with the result given by
(\ref{H_TB_lin}), which can be rewritten in the scaling form (\ref{H_scaling}).
A significant difference at the level of the linearized description 
is the presence of external boundary.
Although the rescaled Hamiltonian $H'_{lin}$ 
contains no information on the system size
$L$ and the microscopic lengthscale $k_F$, the spectrum 
still does not exhibit the scaling 
\begin{equation}
E_{n} = \frac{v_F}{l} F_n(\alpha_{\Delta})~,
\label{scaling0}
\end{equation}
where $F_n$ is a universal function of the anisotropy $\alpha_{\Delta}$, as
one might initially suspect. Instead, both 
the boundary conditions at the grain's edge and the
singular character of the wavefunctions near the vortex affect the
leading order result for the spectrum of the $H_{lin}$, and consequently
violate the simple scaling relation (\ref{scaling0}), whose more
appropriate form should be
\begin{equation}
E_{n} = \frac{v_F}{l} F_n(\alpha_{\Delta}, {\rm B.C.})~.
\label{scaling1}
\end{equation}
The label B.C. here stands for boundary conditions determined by dimensionless
combination $(\tmpQ L)$, that naively might have seemed to drop out of the
leading part of the scaling.

As an illustration, consider first a simple example of a zero-field problem
--  a lattice $d$-wave superconductor in an empty box with impenetrable
walls. The eigenfunctions of the non-linearized problem are given by
\begin{equation}
\psi_{k_x,k_y}(\br) = C \sin (k_x x) \sin(k_y y)
\begin{pmatrix}
u_{\bk}\\
v_{\bk}
\end{pmatrix}~.
\label{zerofieldwf}
\end{equation}
where $C$ is a normalization constant, and due to the zero boundary
conditions at the edges, we have  $k_i =\pi n_i/L$ with positive integer
$n_x$ and $n_y$.

The components of the Nambu spinor $u_{\bk}$ and $v_{\bk}$ can be expressed as
$$
\begin{aligned}
u_{\bk}^2 = (1+\xi_{\bk}/E_{\bk})/2~,\\
v_{\bk}^2 = (1-\xi_{\bk}/E_{\bk})/2~,
\end{aligned}
$$
where $E_{\bk}^2 = \xi_{\bk}^2+\Delta_{\bk}^2$.
\begin{figure}
\includegraphics[width=\columnwidth,clip]{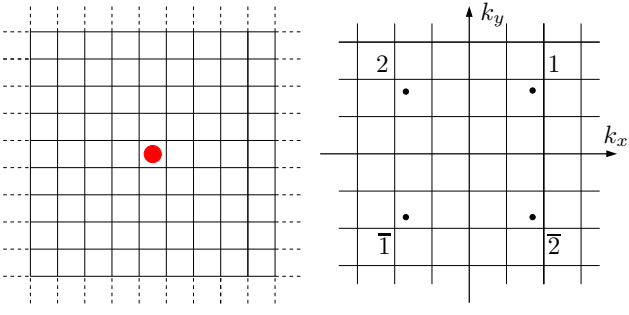}
\caption{\label{combined1}
Left: Superconducting island of size $L\times L$. The
tight-binding lattice is shown in black, the red circle is the vortex, and
the dashed bonds correspond to the boundary of the superconducting region
with  both $t_{\br\br'}$ and $\Delta_{\br\br'}=0$ across the dashed bonds.
solid lines.
Right: Four nodes of a $d$-wave superconductor in momentum
space at $(\pm {\tmpQ}, \pm {\tmpQ})$, where ${\tmpQ}=\delta^{-1}\arccos(-\mu/(4t))$, 
for a general incommensurate  case are shown as four circles. The horizontal and vertical
lines display the grid of
commensurate wavevectors $(\pi n_x/L, \pi n_y/L)$.}
\end{figure}
Note that the solution (\ref{zerofieldwf}) mixes four plane waves
$\exp(i\bk\cdot \br)$ with $\bk=(\pm k_x, \pm k_y)$ in a
specific combination, and no other combinations are allowed. Among other
things, it suggests that simply taking eigenfunctions of the linearized
Hamiltonians corresponding to energy $E$ and combining them in all possible
combinations in $\psi^{(j)}$ (\ref{linearizedWF})  will not work: only  special
superpositions, which make the full wavefunction vanish at the edges of the
system, are allowed. Consider now the lowest energy levels: if the node 1
situated at $({\tmpQ}, {\tmpQ})$, coincides with one of the
allowed mesh points in momentum space $(\pi n_x/L, \pi n_y/L)$, then the
lowest energy value is simply zero. Otherwise, depending on anisotropy
$\alpha_D$ the lowest energy level is reached at one of the four points of
the mesh closest to the node $({\tmpQ}, {\tmpQ})$. 
More precisely, if ${\tmpQ}=\pi n
/L+\delta k$ with $|\delta k|< \pi/(2L)$, then the ground
state energy is given by the least of 
$E_{\pi n/L, \pi n/L}$ and  $E_{\pi n/L,
\pi(n+\sign(\delta k))/L}$,  which to the leading order in $1/L$ equal
\begin{equation}
E_{\pi n/L, \pi n/L} = 4|t\sin ({\tmpQ} \delta) (\delta k)|\label{Enn}\\
\end{equation}
and
\begin{multline}
E_{\pi n/L, \pi(n+1)/L} =\\
2|\sin ({\tmpQ} \delta)| \sqrt{ t^2
\left(2|\delta
k|-\frac{\pi}{L}\right)^2+\tmpDelta^2\left(\frac{\pi}{L}\right)^2}\label{Ennplus1}~.
\end{multline}
Therefore, the ground state energy is given by (\ref{Enn})
when $|\delta k|<(1+\tmpDelta^2/t^2)\pi/(4L)$, and by (\ref{Ennplus1}) otherwise.
Note that the result is  an oscillating function of ${\tmpQ}$ changing from zero,
whenever ${\tmpQ} L = \pi n$ and the nodes  coincide with mesh points, to
$$
 2\pi \min(t,\tmpDelta)\sin({\tmpQ}\delta)/L = \min(t,\tmpDelta)
\frac{2\pi}{L}\sqrt{1-\frac{\mu^2}{16t^2}}~.
$$ 
As a result, strictly speaking, there
is no uniform scaling (\ref{scaling0}) of Simon-Lee type even for the
zero-field  problem, no matter how large the magnetic length $L$ is. Instead,
the scaling is fulfilled only on average.

Why do the linearization, and the scaling relation (\ref{scaling0}) fail? The
answer is that in this problem the Fermi momentum
scale $k_F$ has not truly been eliminated from
the linearized problem. Although the linearized Hamiltonian $H_{lin}'$ does not
contain any dependence on $k_F$, the boundary conditions that should be
satisfied by the eigenfunctions of 
the Hamiltonian {\em retain} the information on
commensuration between $\tmpQ$ and $1/L$. To derive them in general, consider
again (\ref{linearizedWF}). At the left boundary $x=0$, and the condition reads
\begin{multline}
e^{i{\tmpQ} y} \Bigl(\psi^{(1)}(0,y)+\psi^{(2)}(0,y)\Bigr)\\
+ e^{-i {\tmpQ} y} \Bigl(\psi^{(\bar{1})}(0,
y)+\psi^{(\bar{2})}(0,y)\Bigr)=0~.
\end{multline}

Since $\psi_i$ vary slowly on the scale of $1/k_F$, we obtain
\begin{align}
\psi^{(1)}(0,y)+\psi^{(2)}(0,y)&=0~,\\
\psi^{(\bar{1})}(0, y)+\psi^{(\bar{2})}(0,y)&=0~.
\label{bcEXT1}
\end{align}
Similarly, the boundary condition at the right edge of the square is 
\begin{multline}
e^{i{\tmpQ} y} \Bigl(e^{i{\tmpQ}L}\psi^{(1)}(L,y)+e^{-i{\tmpQ}L}\psi^{(2)}(L,y)\Bigr)\\
+e^{-i{\tmpQ} y}
\Bigl(e^{-i{\tmpQ}L}\psi^{(\bar{1})}(L, y)+e^{i{\tmpQ}L}\psi^{(\bar{2})}(L,y)\Bigr)=0~,
\end{multline}
and consequently
\begin{align}
e^{i{\tmpQ}L}\psi^{(1)}(L,y)+e^{-i{\tmpQ}L}\psi^{(2)}(L,y)&=0\\
e^{-i{\tmpQ}L}\psi^{(\bar{1})}(L, y)+e^{i{\tmpQ}L}\psi^{(\bar{2})}(L,y)&=0~.
\label{bcEXT2}
\end{align}
In addition, similar conditions must be satisfied at $y=0$ and at $y=L$.
Note that the conditions on the eigenfunctions of the linearized problem
couple different nodes, and moreover they explicitly involve a phase factor
$\exp(2i{\tmpQ}L)$. It is expected, therefore, that the spectrum of the
problem does depend on ${\tmpQ} L$ modulo $\pi$ even in the leading order
of approximation. 

\begin{figure*}
\includegraphics[width=0.99\textwidth]{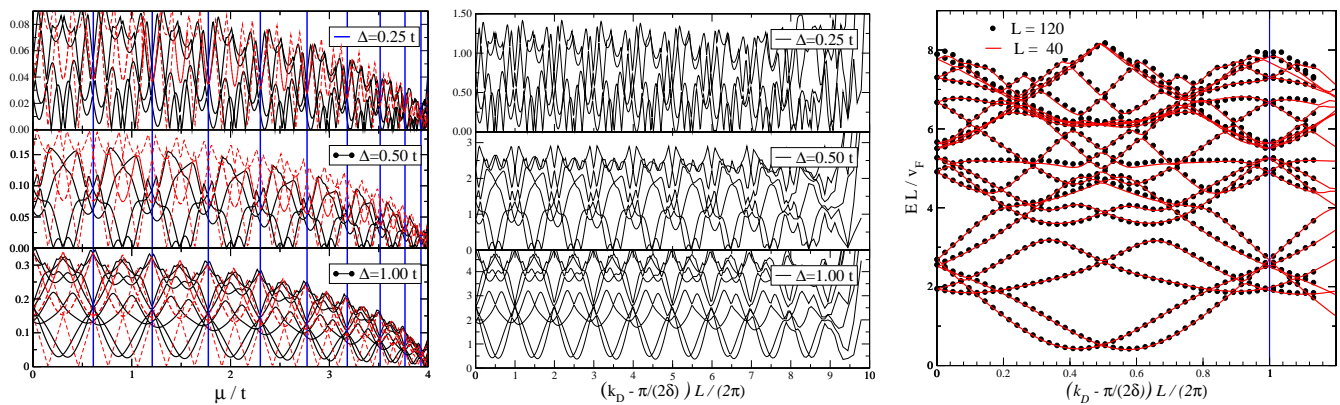}
\caption{\label{fig:spectrum} Left panel: The energy of the lowest $20$ energy levels as a
function of $\mu$ is shown in black solid circles. The three panes of the figure correspond to $\Delta=t/4$,
$\Delta=t/2$, and $\Delta=t$. In all cases, the size of the system $L$
equals $41$. The red dashed lines are the energies of the lowest four
eigenstates  of the zero-field problem.
Center panel: Rescaled version of the left panel. The rescaled energy  
$E_n L/(\hbar v_F)$ levels are plotted versus
 $({\tmpQ}-\pi/(2\delta))L/2\pi$. The $\pi/(2\delta)$ offset is artificially introduced to
make the plots with different $L$ collapse on the same graph near half filling $\mu=0$.
Right: The energy eigenvalues rescaled as in the center panel are shown for
$L=41\delta$ and $L=121\delta$.
Only the first cycle of the oscillatory pattern is shown for compactness. In
both cases, $\Delta=t$.
}
\end{figure*}

The vortex problem adds a new layer of complexity when the
linearization is performed. Although in the full, non-linearized problem,
the divergence is cut-off at about vortex core radius
$\xi$, and does not pose complications, the  effects of the curvature terms such as the inter-nodal
interference are enhanced due to the singular character of the vortex potential
at the  linearized Hamiltonian level, as explained in the previous section.
A simple estimate similar to 
(\ref{curvatureINTER}) shows that as a result, the oscillations of
the energy levels are  controlled by condition
(\ref{theoretical_periodicity}), and thereby the periodicity of the energy
levels in a grain with a vortex,
plotted as a function of ${\tmpQ}L$, should  be doubled compared
to the oscillations in an empty grain.

The spectrum of a tight-binding lattice superconductor of size $L\times L$ 
found numerically in a presence of a single vortex in 
a magnetic field $HL^2=hc/(2e)$,
is shown in Fig. \ref{fig:spectrum} (left panel).
The low-energy quasiparticle spectrum exhibits the following properties:
(i) a generic spectrum is gapped, (ii) for $\tmpDelta<\Delta_c$, where
$\Delta_c\approx 0.75t$, the
zero energy states appear at discrete values of 
the chemical potential $\mu$, (iii) the
spectrum {\em does not} follow a simple scaling relation (\ref{scaling0}) 
displaying instead an oscillatory behavior, with the magnitude of the
oscillations scaling as $\hbar v_F/L$, and the period of oscillations
decreasing away from $\mu=0$. The last property is 
a direct consequence of the approximate $2\pi$-periodicity of
the spectrum with respect to ${\tmpQ}L$: the  
spectra therefore must be similar at
$\mu$ and $\mu'$ related by $L ({\tmpQ}(\mu)-{\tmpQ}(\mu'))=2\pi n$, where
${\tmpQ}(\mu) = \arccos(-\mu/(4t)$, and therefore the 
periodicity condition for
$l\gg \delta$, when the equivalent values 
of $\mu$ are closely spaced,  can be written as 
$$
\delta\mu \approx \frac{2\pi}{L} \sqrt{1-\frac{\mu^2}{16t^2}}~.
$$
Just as in the case of a vortex lattice, it is useful to redisplay the data by
extracting the analogue of the overall Simon-Lee
 scaling factor $ v_F/L$ from the energy levels, by  
plotting $E L/v_F$ vs ${\tmpQ}L$ 
(see Fig. \ref{fig:spectrum}, center).
Clearly, after
such rescaling, the dependence of the energy
levels on $\mu$ (or ${\tmpQ}$) is more 
uniform compared to the raw data plotted in Fig.
\ref{fig:spectrum}. The oscillatory part of the spectrum, however, also
scales as $1/L$ (see Fig. \ref{fig:spectrum}, right panel). 
Note that the periodicity of the
oscillations in  ${\tmpQ} L$ is $2\pi$, {\em twice} the periodicity of the
zero-field problem, as expected. The pattern of Fig. \ref{fig:spectrum} holds
extremely well for all  $\mu$, except near the extreme values close to
$\mu\approx -4t$, where the $k_F$ becomes comparable with  $1/\delta$, and the
linearization procedure is not justified.
The commensuration effects could only be experimentally
accessible at the temperatures smaller than the
amplitude  of the oscillations in energy 
($\approx \hbar v_F /L$ for the lowest energy levels)
The thermally broadened quantities such as the TLDOS
will not reflect these oscillations unless the
 temperature $T  \lesssim \hbar v_F/L$.  Besides the temperature,
impurities, instrumental resolution and 
other factors could result in broadening
the energy levels. 

\begin{figure}
\includegraphics[width=\columnwidth,clip]{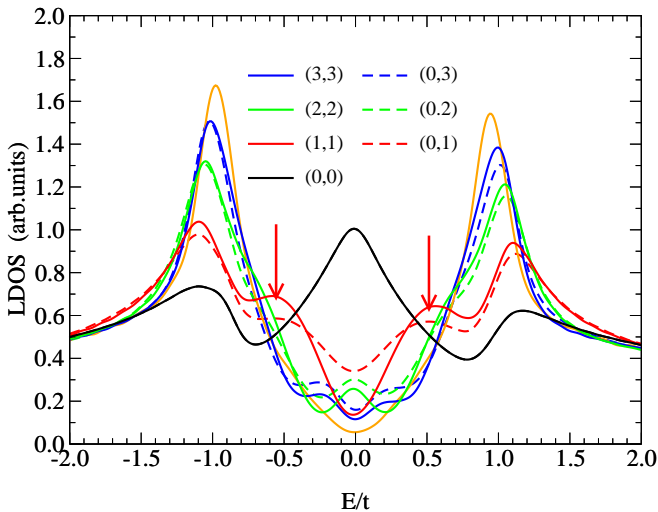}
\caption{\label{fig:singleldos} TLDOS for several sites surrounding the vortex are
shown in black, red, green, and blue. The orange line represents TLDOS
half-way from the vortex and the boundaries of the system. The black curve
with a peak at $E=0$ represents  TLDOS at four sites surrounding the
plaquette with the vortex. The parameters used in this figure are
$L=80\delta$, $\mu=0.2t$, $v_{\Delta} = 0.25 v_F$;  the
main structure  of the graph is robust under the change of parameters of
the model: in particular, at all cites near the vortex, except the four
nearest neighbors, the TLDOS has additional peaks at energy  $\sim
\Delta_c/3-\Delta_c/2$.
}
\end{figure}

Now we turn to large-energy, short-distance
  features of the quasiparticle spectrum and
describe the TLDOS calculated at 
temperatures larger than the separation between
the energy levels, but still smaller than $\Delta$. This implies
sufficiently large $L$. In practice we used typically
$T=0.05t$ and $L$ ranging from $30\delta$ 
to $120\delta$. A representative TLDOS is
shown in Fig. \ref{fig:singleldos}. 
The ${\tmpQ}l$-oscillations at the 
lowest energy scales $C(\alpha_{\Delta})\hbar v_{F}/L$ are
essentially absent in such thermally broadened LDOS.
While the resulting  TLDOS will still  depends on $\mu$ (or ${\tmpQ} l$),
the dependence is not oscillatory and merely reflects the
slow varying changes of the normal state band structure, yielding
large-scale changes such as the position of the van Hove peak. 
Overall, the spacial and energy distribution of TLDOS is quite similar to the
vortex lattice case:
 the TLDOS at the center of the vortex has a zero-bias coherence peak,
 while far from a vortex and the edges the TLDOS is similar to the uniform
 zero-field result, as expected. At the 4 nearest and 4 next nearest neighbors the TLDOS has
pronounced peaks at $1/3-1/2$ of the coherence peak energy. 

\section{Conclusions}
\label{section:conclusions}
We analyzed the properties of the mixed state in the tight-binding lattice
$d$-wave superconductors by considering a quasiparticle spectrum of i) a vortex
lattice and, separately, of  ii) an isolated superconducting grain
accommodating precisely a single
vortex. To reduce the set of free parameters to a manageable number, 
we restricted ourselves to the 
simplest tight-binding model, described by only two
parameters associated with the normal state $(t, \mu)$, and additionally
assume that the magnitude
of the gap function $\tmpDelta$ is spatially uniform. 
Within a set of simple mean-field microscopic theories, such as those
arising from an
extended Hubbard or a $t-J$ model, this assumption 
is well justified since for the
parameters suited for to the  cuprates, with 
their short coherence lengths, the
self-consistent calculations show that the amplitude of the order parameter
recovers its bulk value at distances of
only a few lattice spacings.

We find that the low-energy properties of the
spectrum are described by Simon-Lee scaling 
only on {\em average}, and that both
the energy dispersion and the (local) density of states experience
oscillations as a function of the chemical potential and a magnetic field. The
magnitude of the oscillations in the energy levels behaves as $l^{-1}$ as 
a function of magnetic length, and therefore it is of the same 
order as the average Simon-Lee part of
the dependence of $E_{n\bk}$ on $l$. We find that in all cases, these
oscillations can be well described by a modified Simon-Lee scaling
(\ref{modsimonlee}), which includes {\em additional} 
$2\pi$-periodic dependence of the energy
levels on $\tmpQ L$.
This modification is shown to be 
a consequence of the diverging solutions of the
``linearized'' Hamiltonian. A careful treatment of these divergencies
is needed\cite{mt_SAE}, lest one 
underestimates the quantitative importance of the inter-nodal and
formally subleading intra-nodal matrix elements of the Hamiltonian. As a
result, the actual expansion parameters of the theory include
$k_F\xi$, rather than only
$k_F l$, and the scaling of the quasiparticle energy spectrum with respect
to magnetic length $l$ is consequently modified.

In addition, we analyzed the large-energy, short-distance features of 
the quasiparticle spectrum. We found that -- apart from the four 
sites surrounding  a vortex where
the thermally broadened TLDOS exhibits zero 
bias conductance peak (ZBCP), in agreement with Refs.
\onlinecite{wang_macdonald,ft_DPLUSID} -- the TLDOS on 
all other sites in the vicinity of a
vortex instead show peaks at sub-gap energies. The energy 
of these sub-gap peaks
does not depend on magnetic field, and is determined only by the parameters
of the band structure and $\tmpDelta$.

Finally, we 
examined how the TLDOS is modified when the amplitude of the $d$-wave
gap function
is varied locally in the vicinity of the vortex core and found
that the suppression of the zero-bias peak corresponds to the enhancement of
the  $d$-wave gap function, which could
arise through the locally enhanced\cite{nunner, bma} effective pairing
interaction constant $g$. Such enhancement might result from the effect of the
impurity atoms  pinning the vortices or from a local distortion of
atomic lattice by a vortex or fluctuations of $d$-wave order parameter.

\acknowledgments
We are  indebted to O. Vafek  for correspondence and 
stimulating discussions. A.M. would also like to thank
B.M. Andersen,  A.V. Balatsky, M. Franz, P. J. Hirschfeld, and T. Nunner for discussions.
This work was supported in part by  the NSF grants DMR-0094981
and DMR-0531159.

\appendix*

\section{Phase factors in the tight-binding Hamiltonian}
\label{appendix_tightphases}

\begin{figure}
\parbox{0.62\columnwidth}{\includegraphics[width=0.6\columnwidth]{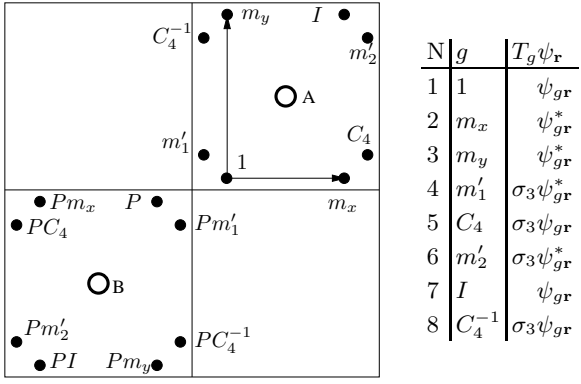}}%
\parbox{0.33\columnwidth}{\begin{tabular}{l|l|l}
 N\strut & $g$         &$T_g\psi_{\br}$\\
 \hline
 1\strut &{$1$}        &$\phantom{\sigma_3}\psi_{g\br}$\\
 2\strut &{$m_x$}      &$\phantom{\sigma_3}\psi^*_{g\br}$\\
 3\strut &{$m_y$}      &$\phantom{\sigma_3}\psi^*_{g\br}$ \\
 4\strut &{$m'_1$}     &$        {\sigma_3}\psi^*_{g\br}$\\
 5\strut &{$C_4$}      &$        {\sigma_3}\psi_{g\br}$\\
 6\strut &{$m'_2$}     &$        {\sigma_3} \psi^*_{g\br}$\\
 7\strut &{$I$}        &$\phantom{\sigma_3}\psi_{g\br}$\\
 8\strut &{$C_4^{-1}$} &$        {\sigma_3}\psi_{g\br}$
 \end{tabular}
}
\caption{\label{sympsi} Left: Point group transformations: equivalent points in
the unit cell, which are obtained by action of the group operation $g$ from
Table \ref{sympsi} on the reference point $1$,  are shown as solid circles. 
Large open circles denote vortices.
Right: The symmetry transformations of  tight-binding  Hamiltonian
${\cal H_{TB}}$. If $\psi_{\br}$ is an  eigenstate
of energy $E$, then the states $\psi'_{\br}$ are also eigenfunctions
of the Hamiltonian with the same energy, but different momenta in
the Brillouin zone listed in the right column of the table. 
For brevity only a half of all transformations is listed. The
remaining half is obtained by applying operator $P\psi_{\br} =
\gamma_{\br}\psi_{-\br}$ to each operation in the Table. Thus, overall
there are 16 symmetry operations requiring that the spectrum is
16-fold symmetric as shown in the right panel of Fig. \ref{unitcell}.}
\end{figure}

As mentioned in the main text, 
in principle the bond phase of the order parameter
$\theta_{\br\br'}$ should be determined self-consistently. It is
convenient, however, to adopt a synthetic approach and approximate
$\theta_{\br\br'}$ by using the known solution $\phi(\br)$
of the {\em continuum} Ginzburg-Landau vortex-lattice problem, whose
explicit form is given for example in Appendix A of Ref. \onlinecite{mt_SAE}.
One may set  $\theta_{\br\br'} =
\phi[(\br+\br')/2]$  or 
$\theta_{\br\br'}=(\phi(\br)+\phi(\br'))/2$. The latter, which will
be predominantly used by us in this article,  requires
explanation since $\theta$ must be defined modulo $2\pi$, while in the form
above $\theta_{\br\br'}$ is defined only modulo $\pi$. More accurate
definition reads
\begin{equation}
\exp(i\theta_{\br\br'})= \exp\left(i \phi(\br)+\frac{i}2 \int_{\br}^{\br'}\nabla
\phi(\br'')\cdot d\br''\right),
\label{definetheta}
\end{equation}
where the integral is over the bond connecting $\br$ and $\br'$. It
is easy to show that this definition is consistent in the sense that 
$\exp(i \theta_{\br\br'})=\exp(i\theta_{\br'\br})$, and provided
that $\phi(\br)$ does not change by more than $\pi$ along a bond,
the phase $\theta_{\br\br'}$ defined in this way is indeed the
``average'' of $\phi(\br)$ and $\phi(\br')$ in the sense that the
``average'' is understood as the closest to either $\phi(\br)$ (or
equivalently $\phi(\br')$) of the two possible choices.

Note that unlike $\theta_{\br\br'}$, phases $\phi^A_{\br}$ and
$\phi_{\br}^B$ are not determined by any self-consistent procedure, and
merely serve as a technical device to recast the Hamiltonian in a periodic
form;  we are therefore free to assign their values according to our
convenience. Without lost of generality, we  choose them by 
simply evaluating continuous functions  $\phi_A(\br)$ and $\phi_B(\br)$
from the previous section on sites of the tight-binding lattice.

Given definition (\ref{definetheta}),  coefficients
$\cV^A_{\br\br'}$, $\cV^B_{\br\br'}$, and
$\cA_{\br\br'}$ can be easily found from (\ref{definebondvars})  and
explicit expressions for $\phi_{A(B)}$ from Appendix A of Ref.
\onlinecite{mt_SAE}:
\begin{align}
e^{i\cV^{\alpha}_{\br\br'}} &= e^{i\int_{\br}^{\br'} \bv_{\alpha}(\br'') \cdot
d\br''}, \quad \alpha=A, B\\
e^{i\cA_{\br\br'}} &= e^{\frac{i}{2}\int_{\br}^{\br'}
(\bv_A(\br'')-\bv_B(\br''))
\cdot d\br''}
\end{align}
Since $\bv_A(\br)$ and $\bv_B(\br)$ are periodic\cite{vmft, mt_SAE}, 
clearly so are $\cV^{\alpha}_{\br\br'}$ and $\cA_{\br\br'}$.

The phase factors possess a number of discrete symmetries, which
result in the symmetry of the dispersion shown in the right panel of
Fig. \ref{unitcell}. The full set of the  symmetry operations consists
of operations $g$ shown in  \ref{sympsi} and  eight additional
transformations $P g$, where $P\psi_{\br}\equiv \gamma_{\br}\psi_{-\br}$
is an inversion operator around a midpoint between two (arbitrary) 
vortices $A$ and $B$. Each transformation involves a point group
operation $g\br$  shown in Fig. \ref{sympsi}, which may be
followed by complex conjugation and multiplication by Pauli matrix 
$\sigma_3$. Thus, overall there are $16$ distinct points in the Brillouin zone
with identical set of energy eigenvalues. 

Additionally, if $\psi_{\br}$ is an eigenstate of energy $E$, then
$\sigma_2\psi^*_{-\br}$ is an eigenstate of energy $(-E)$ and the
same momentum $\bk$. Thus, at each point the spectrum is symmetric
as a function of energy.


\end{document}